\pdfoutput=1
\documentclass[fleqn,usenatbib]{mnras}

\usepackage{newtxtext,newtxmath}
\usepackage[T1]{fontenc}
\usepackage{ae,aecompl}

\usepackage{amsmath}
\usepackage{natbib}
\usepackage{hyperref}
\usepackage{xcolor}
\usepackage{graphicx}
\usepackage{chngcntr}
\usepackage{rotating}
\usepackage{threeparttable}
\usepackage{xcolor}

\newcommand{\alm}{A_{\lambda, \mu}}
\newcommand{\myemail}{giovanni.bruno@inaf.it}
\newcommand{\Ispot}{I_\bullet}
\newcommand{\Istar}{I_\star}
\def\gtsima{$\; \buildrel > \over \sim \;$}
\def\ltsima{$\; \buildrel < \over \sim \;$}
\def\gtrsim{\lower.5ex\hbox{\gtsima}}
\def\lesssim{\lower.5ex\hbox{\ltsima}}

\def\rj{$R_\mathrm{J}$}
\def\rs{$R_\odot$}

\title{Hiding in plain sight: observing planet-starspot crossings with the \textit{James Webb Space Telescope}}

\author[G. Bruno et al.]{
Giovanni Bruno,$^{1}$\thanks{E-mail: \myemail}
Nikole K. Lewis,$^{2}$
Jeff A. Valenti,$^{3}$
Isabella Pagano,$^{1}$
\newauthor
Tom J. Wilson,$^{4}$
Everett Schlawin,$^{5}$
Joshua Lothringer,$^{6,7}$
Antonino F. Lanza,$^{1}$
\newauthor
Jonathan Fraine,$^{8,9}$
Gaetano Scandariato,$^{1}$
Giuseppina Micela,$^{10}$
Gianluca Cracchiolo$^{11,10}$
\\
$^{1}$INAF -- Catania Astrophysical Observatory, Via Santa Sofia, 78, 95123, Catania, Italy\\
$^{2}$Department of Astronomy and Carl Sagan Institute, Cornell University, 122 Sciences Drive, Ithaca, NY 14853, USA\\
$^{3}$Space Telescope Science Institute, 3700 San Martin Drive, Baltimore, MD 21218, USA\\
$^{4}$University of Exeter, Physics Building, Stocker Road, Exeter EX4 4QL, UK\\
$^{5}$University of Arizona, 933 Cherry Ave, Tucson, AZ 85721, USA\\
$^{6}$Department of Physics and Astronomy, Johns Hopkins University, Baltimore, MD 21210, USA\\
$^{7}$Department of Physics, Utah Valley University, 800 W. University Parkway, MS 179, Orem, UT 84058, USA\\
$^{8}$Space Science Institute Center for Data Science, 4765 Walnut St., Suite B, Boulder, CO 80301, USA\\
$^{9}$Space Science Institute Center for Exoplanet and Planetary Science, 4765 Walnut St., Suite B, Boulder, CO 80301, USA\\
$^{10}$INAF -- Osservatorio Astronomico di Palermo, P.za Parlamento 1, 90134 Palermo, Italy\\
$^{11}$Dipartimento di Fisica e Chimica Emilio Segré, Università di Palermo, Via Archirafi 36, 90123 Palermo, Italy
}

\date{Accepted XXX. Received YYY; in original form ZZZ}

\pubyear{2020}

\begin{document}
\label{firstpage}
\pagerange{\pageref{firstpage}--\pageref{lastpage}}
\maketitle

\begin{abstract}
Transiting exoplanets orbiting active stars frequently occult starspots and faculae on the visible stellar disc. Such occultations are often rejected from spectrophotometric transits, as it is assumed they do not contain relevant information for the study of exoplanet atmopsheres.
However, they can provide useful constraints to retrieve the temperature of active features and their effect on transmission spectra. We analyse the capabilities of the \textit{James Webb Space Telescope} in the determination of the spectra of occulted starspots, despite its lack of optical wavelength instruments on board. Focusing on K and M spectral types, we simulate starspots with different temperatures and in different locations of the stellar disc, and find that starspot temperatures can be determined to within a few hundred kelvins using NIRSpec/Prism and the proposed NIRCam/F150W2$+$F322W2's broad wavelength capabilities. Our results are particularly promising in the case of K and M dwarfs of mag$_K \la 12.5$ with large temperature contrasts.
\end{abstract}


%%%%%%%%%%%%%%%%%%%%%%%%%%%%%%%%%%%%%%%%%%%%%%%%%%

%%%%%%%%%%%%%%%%% BODY OF PAPER %%%%%%%%%%%%%%%%%%

\section{Introduction}\label{intro}

The contamination from the stellar signal in transit spectrophotometry is one of the main challenges to precisely characterise exoplanet atmospheres. Stellar active features, such as dark starspots and bright faculae, both hamper the correct measure of transit parameters and introduce spurious spectral features that overlap with the exoplanetary spectral lines. A large body of observational evidence is now available about general starspot properties as a function of stellar type \citep[e.g.][]{berdyugina2005,strassmeier2009,balona2016,savanov2019,nielsen2019,savanov2019,herbst2021}, and important progress in the theoretical starspot modelling has been made \citep[e.g.][]{yadav2015,panja2020}. Predicting the appearance and properties of stellar active features remains however elusive and, in the interpretation of an exoplanet observation which is potentially contaminated by stellar activity, the statistical knowledge of starspots and faculae only provides with partial guidance. For example, different starspot sizes and temperatures have been shown to have a dramatically different impact on transmission spectra \citep[e.g.][and references therein]{zellem2017,rackham2018,rackham2019}, and for individual stars we might not be able to recognise the most likely scenario without additional observational constraints (such as e.g. photometric monitoring, \citealp{huitson2013}). 
 
Moreover, the impact of stellar active features in spectrophotometry is at least twofold, according to whether they lie on the part of the stellar disc intersected by the transit chord or not. The so-called ``non-occulted'' features affect several transit parameters and cannot be easily identified with single-transit information alone \citep[e.g.][]{alonso2008,pont2008,pont2013,czesla2009,leger2009,silva-valio2011,sing2011,ballerini2012,csizmadia2013,barros2013,mccullough2014,oshagh2014,barstow2015,zellem2017,rackham2017,rackham2018,rackham2019,alam2018,wakeford2019}. For this reason, long-duration photometric observations of the star need to be secured around the transits, in order to monitor its activity level. One downfall of this method is, however, that the coverage fraction of the stellar disc in active features and their temperature are strongly correlated parameters: usually, the starspot temperature is assumed in order to extract a wavelength-dependent correction factor for the transit depth, or vice-versa \citep[e.g.][]{alam2018}. Simultaneous multi-wavelength observations were shown to be promising approaches to break this degeneracy \citep[e.g.][]{rosich2020,cracchiolo2021}.

``Occulted'' active features have an opposite effect on the transit depth compared to non-occulted features, and can often be easily identified in the transit profile through a typical flux bump (or dip, depending on whether they are starspots or faculae). They can therefore be modelled \citep[e.g.][]{huber2010,sing2011,pont2013,fraine2014,montalto2014,beky2014,tregloan-reed2015,bruno2016,scandariato2017,mancini2017,louden2017,espinoza2017} and provide useful constraints to characterise both the planet and the host star. With a good enough signal-to-noise ratio (SNR), their temperature, size, and coordinates can be determined, even if a certain degree of degeneracy is unavoidable \citep{beky2014,scandariato2017,mancini2017,morris2017}. In combination with out-of-transit photometric observations, occulted starspots can provide some indications on the temperature and size of the active features on the stellar disc, together with some priors on their filling factor. This attractive possibility can be applied when the SNR of the occultation is large, while the impact of occulted starspots and faculae becomes more subtle when they are so small and tightly grouped that they do not stand out against the transit profile. In this case, they can potentially bias the determination of the transit bottom altogether in a similar, even if opposite, way to non-occulted features \citep[e.g.][]{czesla2009,ballerini2012}. The impact of such an unresolved background of active features, similarly to the one of stellar granulation, could be particularly relevant for terrestrial planets orbiting Sun-like stars \citep{sarkar2018}.

The effect of starspots and faculae on transmission spectroscopy is stronger in the visible, as that is where their brightness contrast with the stellar photosphere is the largest. For this reason photometry, as well as spectroscopy in the visible, were successfully used both from the ground and from space to constrain some of the properties of the occulted active features accidentally detected during transit observations \citep[e.g.][]{silva-valio2010,sing2011,beky2014,mancini2017}. 

As the starspot brightness contrast is lower in the near-infrared (NIR), the occultation signal often falls below the noise level, or however at a level where it does not enable the robust determination of starspot parameters \citep[e.g.][]{bruno2018_w52}. For FGK stars, it is generally assumed that both the impact of occulted and non-occulted active features is negligible in the NIR \citep[e.g.][]{zellem2017,rackham2019}, while the contamination spectrum was shown to be much more problematic for M dwarfs \citep{rackham2018,iyer2020}. At the same time, active features produce small but measurable variations in the limb darkening coefficients, which can be measured in this spectral region \citep{morello2017}. 

Exoplanet searches are increasingly focusing on small, cool stars, as they are favourable targets for the detection and characterisation of low-mass planets in the habitable zone; we expect many more interesting targets than the ones already known to be revealed by the analysis of the \textit{Transiting Exoplanet Survey Satellite} (\textit{TESS}) data \citep{ricker2014}. For this reason, as well as the imminent launch of the \textit{James Webb Space Telescope} (\textit{JWST}), it is crucial to identify realistic strategies to determine the properties of active features and their level of spectral contamination for exoplanets orbiting low-mass stars. Moreover, high-precision observations of starspot and facula crossings could also provide important constraints for planets orbiting warmer stars \citep[e.g.][]{beaulieu2008,desert2011_189,bruno2020}, which will become the focus of future searches for long-period, terrestrial planets (such as \textit{PLATO}, \citealp{rauer2014}). Such information would come for free from transmission spectroscopy observations, and would be valuable both for the purpose of exoplanet characterisation and to better understand stellar activity itself.

In this work, we study the possibilities opened up by \textit{JWST} to constrain the temperature of occulted active features on K and M stars in several geometric configurations and activity levels. We focus on dark starspots, assuming that the same line of reasoning we use here can be adopted for bright faculae. Section \ref{starspotspectrum} describes our working framework, and in Section \ref{simulations} we present our simulations; the results of our analysis are detailed in Section \ref{res}. Section \ref{disc} is dedicated to the conclusions that can be drawn from this work, as well as to future perspectives.

\section{Contrast ratio of active features}\label{starspotspectrum}
The brightness contrast between an active feature and the stellar photosphere, $\alm$, is a function of wavelength $\lambda$ and $\mu \equiv \cos \theta$, where $\theta$ is the angle between the normal to the stellar surface and the line of sight. It can be expressed as \citep[e.g.][]{silva2003}
\begin{equation}
    \alm \equiv 1 - \frac{\Ispot(\lambda, \mu)}{\Istar(\lambda, \mu)},
    \label{planck}
\end{equation}
where $\Ispot(\lambda, \mu)$ and $\Istar(\lambda, \mu)$ are the active feature and stellar specific intensity, respectively. In this formulation, $\alm = 1$ denotes maximum contrast -- a completely dark spot -- and $\alm = 0$ no contrast at all -- feature and stellar photosphere are indistinguishable. A negative contrast indicates a warmer structure, such as a facula, which is brighter than the stellar photosphere. 

The intensity of active features can be represented by specific intensity model spectra with a different effective temperature than the one describing the host star. In addition, starspots are thought to be characterised by a 0.5-1 dex lower $\log g$ than the stellar one, because of the decrease in gas pressure caused by increased magnetic pressure in the darker photospheric upper layers \citep[e.g.][]{solanki2003}. 

\begin{figure*}
\includegraphics[width=\columnwidth]{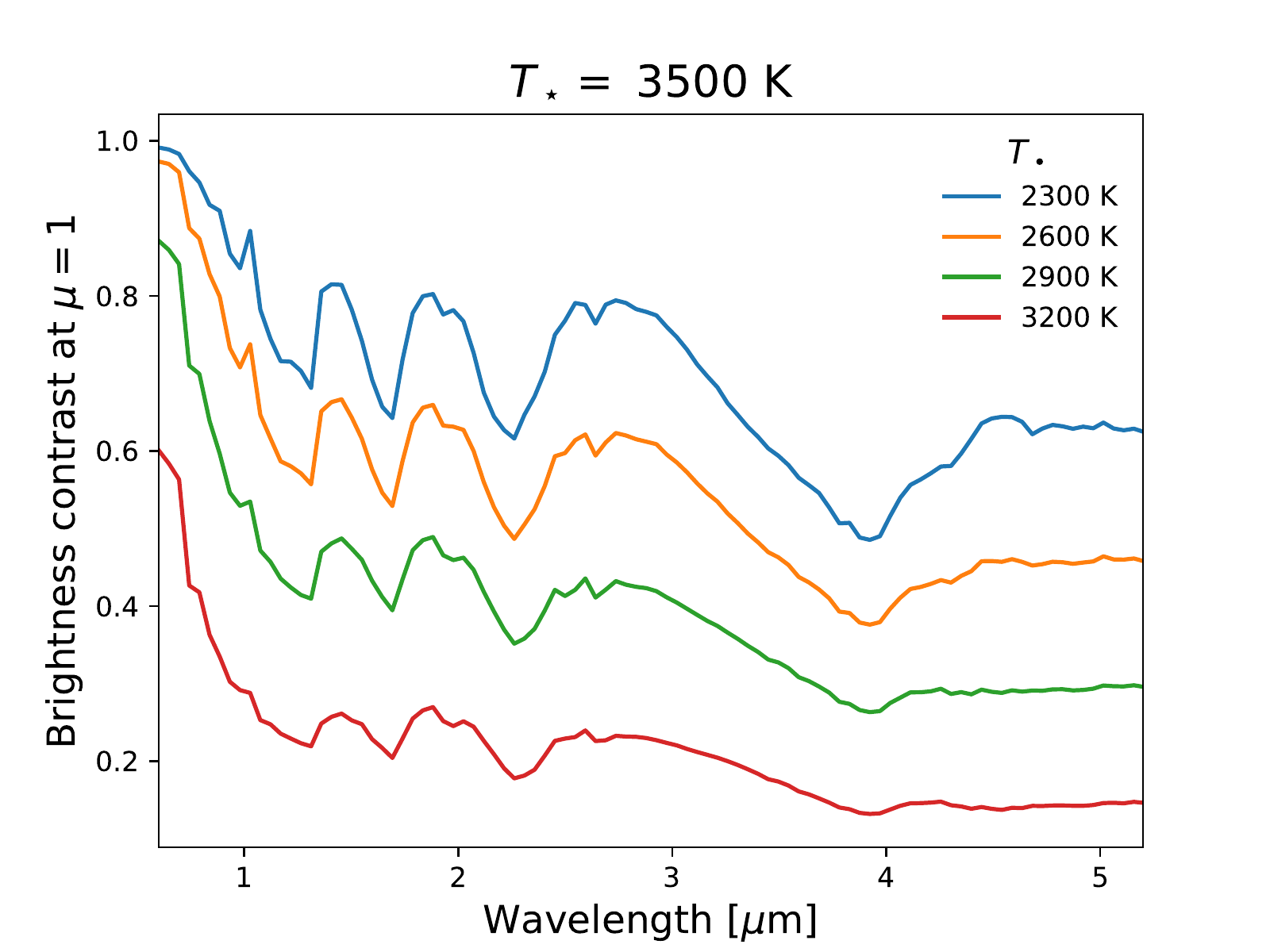}
\includegraphics[width=\columnwidth]{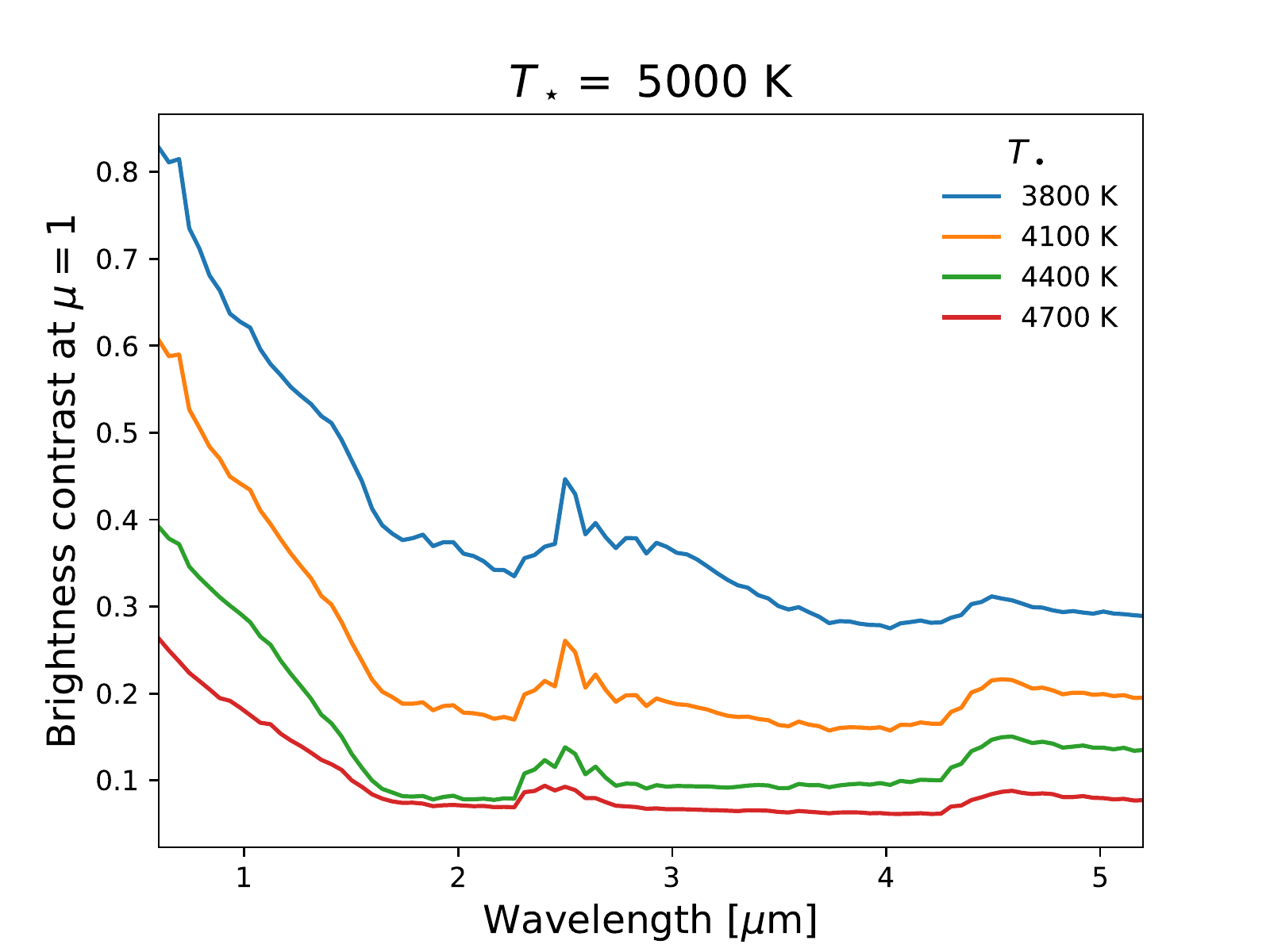}
\caption{Starspot vs. stellar photosphere brightness contrast ratios at the centre of the stellar disc ($\mu=1$), as a function of wavelength, for different stars and starspot effective temperatures (Equation \ref{planck}). The contrast ratios were obtained with PHOENIX specific intensity models and downgraded to resolution $R=100$. Different starspot effective temperatures are represented with different colours, as indicated in the top-right corner of each panel. \textit{Left:} Case of a 3500~K, $\log g = 5.0$, [Fe/H]~$=0.0$ star and starspots with varying $T_\bullet$, $\log g = 4.5$ and the same metallicity. \textit{Right:} Case of a 5000~K, $\log g = 4.5$, [Fe/H]~$=0.0$ star, and starspots with varying $T_\bullet$, $\log g = 4.0$ and the same metallicity. Water vapour bands are visible for the 3500~K star, while CO and OH bands can be noticed for the K star.}
\label{contrasts}
\end{figure*}

As both $\Ispot$ and $\Istar$ depend on $\mu$, the effect of limb darkening (LD) needs to be included in the calculation of the contrast \citep[e.g.][]{dorren1987}. Given this and the complications in describing the limb-brightening behaviour of faculae \citep[e.g.][]{spruit1976,keller2004,solovev2019}, in our analysis we focused on dark starspots. The limb-angle dependent specific intensity stellar models were computed with the PHOENIX stellar atmosphere models \citep{hauschildt1999}. We calculated a baseline model for both the M-dwarf ($T_\star=3500$~K, $\log g=5.0$) and K-dwarf ($T_\star = 5000$~K, $\log g = 4.5$), and then computed models for the starspots every 100 K between 2300 K to 3400 K for the M-dwarf and between 3600 K and 4900 K for the K-dwarf. The M-dwarf starspot models had a $\log g = 4.5$ and the K-dwarf starspot models had $\log g = 4.0$. All models assumed solar metallicity. For each model, once the temperature structure was converged, we calculated the specific intensity through the atmosphere at 51 different $\mu$ values between 0 and 1 at every \AA ngstrom between 6000 and 5.35 \AA.

Figure \ref{contrasts} presents a set of starspot contrast spectra for a $T_\star=3500$~K and a $T_\star=5000$~K star with increasingly cooler spots, calculated from PHOENIX stellar models. The spectra were downgraded to a resolution $R=100$, considering the NIRSpec/Prism's resolution in the spectral range $0.6-5.3 \, \mu$m.
Below $\simeq 4 \, \mu$m, several absorption features, mainly due to water vapour, increase in strength as the temperature contrast between the starspot and the rest of the stellar photosphere increases. Molecular water bands can indeed be observed in stars with spots cooler than $\sim 3000$~K \citep[e.g.][]{wallace1995,jones2002}, so that they can be used as an indicator of the temperature of active features. For the K star, CO absorption at 2.3 and 4.5 microns, and OH absorption at 2.5 microns, also increase with increasing temperature contrast. While the TiO spectral lines at 567, 705.5 and 886 nm are also known as robust temperature indicators \citep[e.g.][]{wing1967,vogt1979,vogt1981,ramsey1980,neff1995,mirtorabi2003,oneal2004,bidaran2016}, a spectral resolution larger than $\sim 10000$ would be needed to achieve significant constraints. Moreover, such lines were only observed in very active stars, where starspots occupy a significant fraction of the visible stellar disc, but which are usually excluded from exoplanet searches.

Starspots that are occulted during transit can be better characterised than those which are not. For example, all latitude values for non-occulted starspots are consistent with observations, other than those spanned by the transit chord; moreover, out-of-transit monitoring with a comparable duration to the stellar rotation period are needed to constrain starspot longitudes. When a starspot is occulted by a planet, its latitude and $\mu$ position on the stellar disc can instead be determined from transit information alone. 

During the occultation of a starspot, the relative flux received from the system increases, creating a ``bump'' in the transit profile. The peak of this bump, $\Delta f$, strongly depends on $T_\bullet$. \cite{sing2011} related $\Delta f$ in the wavelength bin centred at $\lambda$ and at position $\mu$ to the bump at a reference wavelength $\lambda_0$ at the same $\mu$ (their Equation 3):
\begin{equation}
    \frac{\Delta f_\lambda}{\Delta f_{\lambda_0}} = \frac{1 - I_{T_\bullet}(\lambda, \mu)/I_{T_\star}(\lambda, \mu)}{1 - I_{T_\bullet}(\lambda_0, \mu)/I_{T_\star}(\lambda_0, \mu)}.
    \label{crelorig}
\end{equation}
In their formulation, the choice of reference wavelength becomes important, and information is lost in the regions of the contrast spectrum that are around $\lambda_0$.

To avoid any assumption on the reference wavelength, we decided to work with absolute quantities. To derive the flux change that is observed during a starspot occultation, we started by computing the flux observed during a transit if the starspot were at the same $\mu$ position but not occulted, e.g. if it were at an opposite stellar latitude with respect to the centre of the transited stellar belt. In this case, we modelled the nominal transit configuration \citep[e.g.][]{mandelagol2002}, and made the $\mu$-dependent stellar specific intensity $I_\star$ explicit, in place of the integrated flux:
\begin{equation}
    F_1 = F_{\star + \bullet} - \frac{dA_\mathrm{p} \mu_\bullet}{R_\star^2} I_\star(\lambda, \mu_\bullet) = F_{\star + \bullet} - \pi \bigg( \frac{R_\mathrm{p}}{R_\star} \bigg)^2
    I_\star(\lambda, \mu_\bullet),
\label{F1}
\end{equation}
where $F_{\star + \bullet}$ denotes the flux emitted by the star when the starspot is visible on its disc before occultation, and $d A_\mathrm{p} \mu_\bullet /R_\star^2$ represents the solid angle intercepted by the planetary disc on the stellar disc. We used the subscript $\mu_\bullet$ to indicate that the limb angle is evaluated at the centre of the starspot, and we dubbed the planetary surface $dA_\mathrm{p}$ to indicate that it is small compared to the stellar surface. This allowed us to assume that the portion of the stellar photosphere covered by the planet can be described by a single $\mu_\bullet$ value \citep{marino1999,ballerini2012}.

When the starspot is occulted, the stellar specific intensity needs to be replaced by the starspot specific intensity,
\begin{equation}
    F_2 = F_{\star + \bullet} - \pi \bigg( \frac{R_\mathrm{p}}{R_\star} \bigg)^2
    [\beta I_\bullet(\lambda, \mu_\bullet) + (1 - \beta) I_\star(\lambda, \mu_\bullet)],
\end{equation}
where $\beta$ represents the fraction of the planetary disc that occults the starspot, and $1 - \beta$ the part that covers the unspotted stellar disc. This formalism allows the description of planets whose projection on the stellar disc is larger than the occulted starspot, as in this case $1 - \beta > 0$.

The difference between these two quantities is
\begin{equation}
    F_2 - F_1 = \pi \beta 
    \bigg( \frac{R_\mathrm{p}}{R_\star} \bigg)^2     [I_\star(\lambda, \mu_\bullet) - I_\bullet(\lambda, \mu_\bullet)].
\label{deltaI}
\end{equation}

To compute the occultation flux bump in normalised quantities, we then divided $F_2 - F_1$ by the spotted stellar flux $F_{\star + \bullet}$. To do this, we noticed that in an observational context one works with the measured planet-to-stellar surface ratio, which is affected by the presence of starspots. True and observed transit depth can be related by a wavelength-dependent factor that multiplies the stellar flux:
\begin{equation}
 \bigg( \frac{R_\mathrm{p}}{R_\star} \bigg)_\mathrm{true}^2 \equiv \bigg( \frac{\Delta F}{F_\star} \bigg) = \alpha(\lambda) \bigg( \frac{\Delta F}{F_{\star + \bullet}} \bigg) \equiv \alpha(\lambda)\bigg( \frac{R_\mathrm{p}}{R_\star} \bigg)_\mathrm{obs}^2,
\end{equation}
where we used the subscripts ``obs'' for the measured transit depth, $F_\star$ is the unspotted stellar flux, $\Delta F$ indicates the stellar flux drop during transit, and $\alpha < 1$ for a dark starspot \citep[e.g.][]{czesla2009}. Given this definition,
\begin{equation}
 \alpha(\lambda) = F_{\star + \bullet}/F_\star.
 \label{alpha}
\end{equation}
By plugging Equation \ref{alpha} in Equation \ref{deltaI}, and dividing by $F_{\star + \bullet}$, the two $F_{\star + \bullet}$ terms conveniently cancel out:
\begin{equation}
 \Delta f (\lambda) \equiv \frac{F_2 - F_1}{F_{\star + \bullet}} = \pi \beta  \bigg( \frac{R_\mathrm{p}}{R_\star} \bigg)_\mathrm{obs}^2 \frac{I_\star(\lambda, \mu_\bullet)  - I_\bullet(\lambda, \mu_\bullet)}{F_\star}.
 \label{deltaf_f}
\end{equation}
This derivation is therefore independent of the presence of additional non-occulted starspots on the stellar disc, as they can be absorbed by the $F_{\star + \bullet}$ parameter, and then removed by the $\alpha(\lambda)$ factor. The last term to be made explicit is the unspotted stellar flux, which was calculated as \citep[e.g.][]{gray1976}
\begin{equation}
       F_{\star} = 2 \pi \int_{0}^{1} I_\star(\lambda, \mu) \mu d\mu. 
       \label{fluxfromi}
\end{equation}

We again remark that this derivation depends on the assumption that the starspot can be represented by a single $I_\bullet(\lambda, \mu_\bullet)$. In reality, large starspots or starspot groups might be better represented by a combination of the $\mu$ values they span. In this regard, the assumptions of our formalism are the weakest when $\mu = 1$: in this case the non-occulted flux $F_1$ cannot be calculated with Equation \ref{F1}. However, this is not a problem if the resolution of the $\mu$-grid allows both the projected planet and the starspot to lie in the stellar annulus at $\mu=1$ without overlapping. This holds true in the hypothesis of small planets and starspots compared to the stellar disc.

The quantity $\Delta f(\lambda)$ can be fitted to the relative measured flux bumps in the transit profile, in order to derive $T_\bullet$ (assuming $T_\star$ is known to a sufficient level of precision) and $\beta$. The $T_\bullet$ and $\beta$ parameters will be therefore prone to the degeneracy between the starspot temperature and size. Moreover, the fraction of the spot's surface which is actually occulted by the planet cannot be constrained from the occultation event alone, but only a lower limit on its size can be placed by the duration of the occultation event,  as we will discuss in Sections \ref{spectrafit} and \ref{res_deg}.

For our simulations, we chose two \textit{JWST} modes that allow for exoplanet transmission spectroscopy observations over the broadest range of near-IR wavelengths. The NIRSpec/Prism mode, in the range $0.6-5.3 \, \mu$m, will offer constraints on starspot and facula contrast spectra for dim targets; also, it will often provide the most of information for planet atmospheres \citep{batalha2017_ic}. This mode would reach saturation on targets brighter than mag$_K \simeq 10.5$: hence, we also performed simulations where the NIRCam's Dispersed Hartmann Sensor (DHS) would enable simultaneous $1.0-2.0 \, \mu$m and $2.4-4.0 \, \mu$m observations on bright targets by combining the F150W2 and the F322W2 filters \citep{schlawin2016}, if this mode becomes operational. 

\section{Simulations}\label{simulations}
The goal of our simulations was to predict the constraints that \textit{JWST} NIRSpec/Prism and NIRCam/F150W2+F322W2 will be able to place on the effective temperature of occulted starspots for stars with different magnitudes, and spanning a wide range of activity levels. For each observing mode, we proceeded as follows:
\begin{enumerate}
    \item We considered two systems, a K and an M star, each hosting a planet with a different radius. We simulated wavelength-dependent uncertainties on the transmission spectra for a number of different stellar magnitudes, and assumed the transmission spectra to be featureless (to represent a thin or cloudy planetary atmosphere).
    \item Given each transmission spectrum, we modelled the spectrophotometric transits for every wavelength bin. We added a starspot occultation to each transit, using the same spot configuration for every point in the transmission spectrum. We repeated the simulation for a range of starspot temperatures, representing different stellar activity levels, as well as different positions on the stellar disc.
    \item For each one of these scenarios (identified by a stellar type, stellar magnitude, starspot temperature and location), we fitted all associated spectrophotometric transits in order to obtain $\Delta f(\lambda)$ (Equation \ref{deltaf_f}). Such measures were then used to sample the contrast spectra of the occulted starspot, and were fitted to stellar specific intensity spectra to determine the distance from the input starspot temperatures $T_\bullet$.
\end{enumerate}

\begin{figure*}
\includegraphics[scale=0.5]{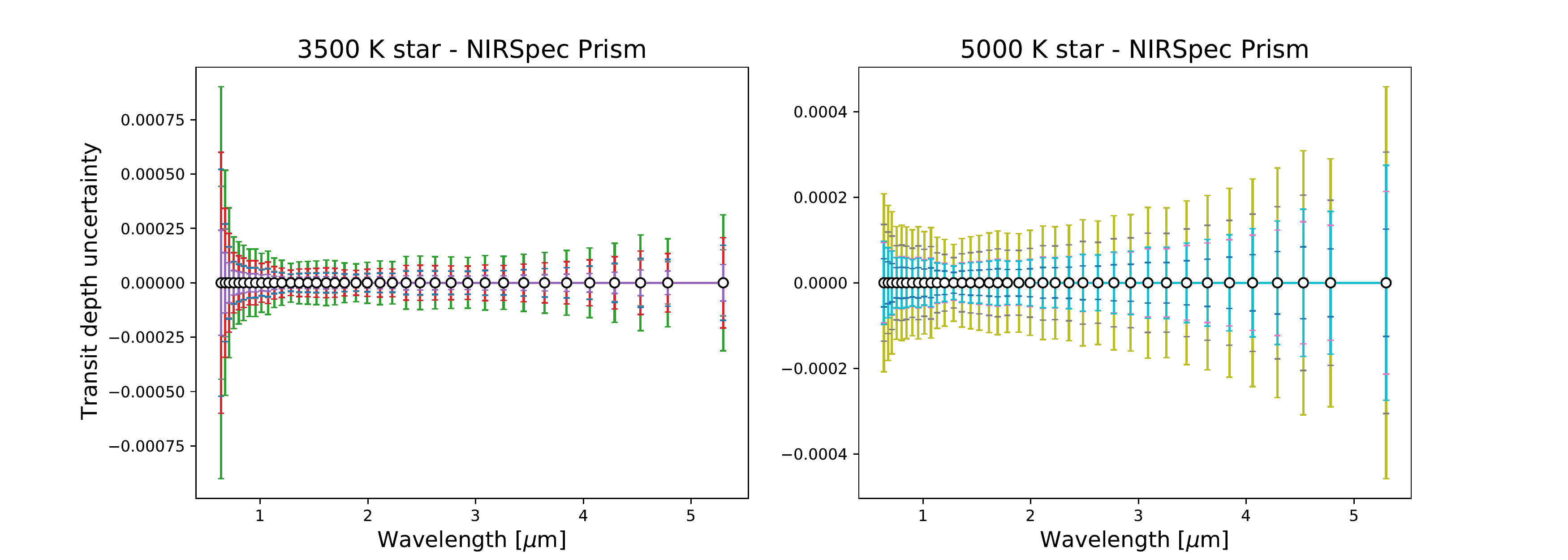}
\includegraphics[scale=0.5]{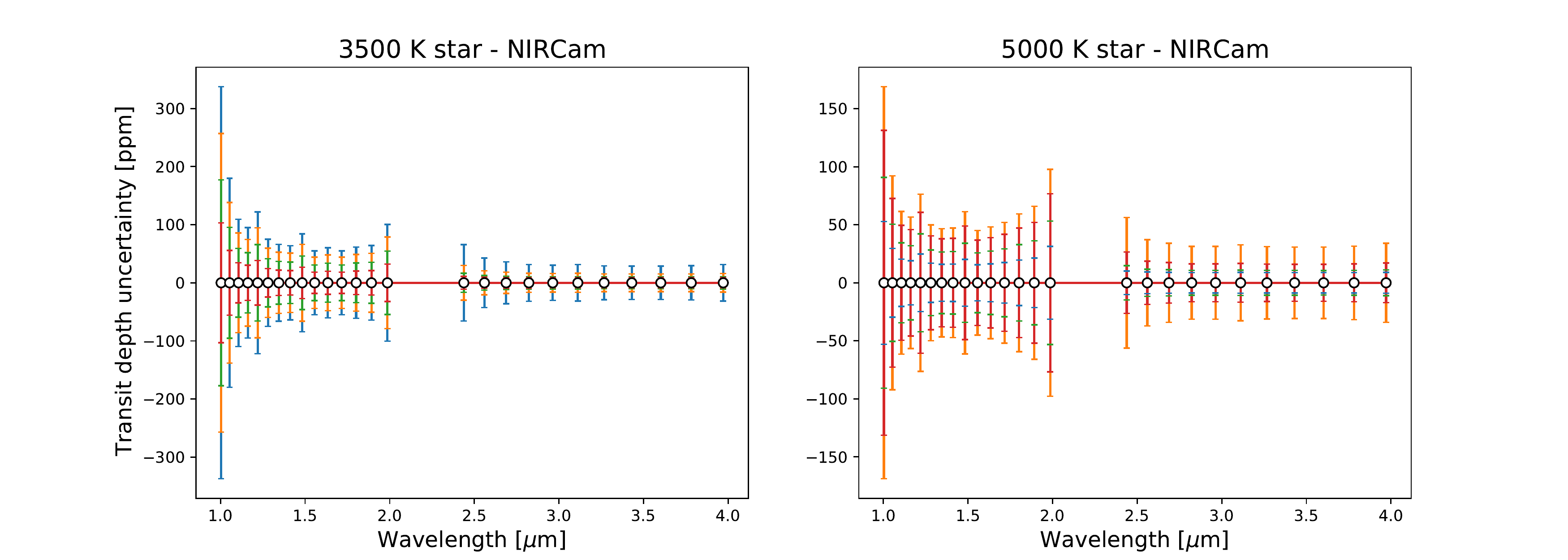}
\caption{Transmission spectra uncertainties computed for the NIRSpec/Prism (\textit{top}) and NIRCam/F150W2 + F322W2 (\textit{bottom}) configurations. Increasingly larger uncertainties correspond, for NIRSpec, to mag $K=10.5, 11.5, 12.5, 13.5$ and 14.5; for NIRCam, they correspond to $K=4.5, 6.0, 7.5$ and 9.0. M and K star scenarios are shown on the left and right column, respectively.}
\label{spec_unc}
\end{figure*}

\subsection{Transit modelling}\label{transitmodelling}
For NIRSpec/Prism, we used the \textsc{PandExo} software \citep{batalha2017} to simulate the transmission spectra uncertainties, using BT-Settl stellar spectral energy distributions. PandExo includes the contribution from shot, background and read noise, with a dependence on the ratio between the duration of out- and in-transit observations, that we set equal to 1. We used the optimisation function in the software that selects the maximum possible number of groups\footnote{This is the number of non-destructive frames that are averaged onboard before being recorded: see https://jwst-docs.stsci.edu/understanding-exposure-times for more detail.} before saturation (fixed to 80\% of the full well), and adopted the NRSRAPID readout pattern. The simulations were run for the SUB512 subarray, and a noise floor of 20 parts per million was adopted. For NIRCam, we used a custom SNR estimator \citep{greene2016,schlawin2016} and PHOENIX spectra. We assumed the maximum number of groups that can be achieved to be 0.5 magnitudes below the saturation limit for the F322W2 grism time series and the RAPID read mode. For the simultaneous short wavelength DHS observations, we adopted the same exposure parameters as the F322W2 grism because the software requires simultaneous readouts. The size of the subarray determines the number of DHS spectra that may be captured as well as the minimum magnitude that will saturate the detector. The balance between capturing DHS spectra while not saturating subarrays results in 1 DHS for $K \leq 6.0$, 2 DHS for $K>6$ and $ \leq 8.3$, and 10 DHS for $K>8.3$. The resulting uncertainties as a function of stellar magnitude and wavelength are shown in Figure \ref{spec_unc}.

\begin{table*}
\caption{Simulated scenarios. From left to right, the columns indicate modelled instrument and observing mode, stellar $K$ mag, stellar effective temperature, stellar surface gravity, starspot simulated effective temperature range and step, planet radius, stellar radius, starspot limb angle $\theta_\bullet$, starspot latitude on the stellar disc $\xi$, and planet orbital inclination $i$.}
\label{tabinput}
\begin{center}
\begin{tabular}{lccccccccc}
\hline \hline
Instrument/mode & $K$ mag range & $T_\star$ [K] & $\log g$ [cgs] & $T_\bullet$ range (step) [K] & $R_\mathrm{p}$ [$R_\mathrm{J}]$ & $R_\star$ [$R_\odot$] & $\theta_\bullet [^\circ$] & $\xi [^\circ]$ & $i [^\circ]$\\
\hline 
NIRSpec/Prism & 10.5-14.5 & 5000 & 4.5 & 3800-4700 (300) & 0.75 & 0.75 & 0, 21, 40 & 0, 21 & 90\\
NIRSpec/Prism & 10.5-14.5 & 3500 & 5.0 & 2600-3200 (300) & 0.25 & 0.47 & 0, 40 & 0 & 90\\
NIRCam/F150W2 + F322W2 & 4.5-9.0 & 5000 & 4.5 & 3800-4700 (300) & 0.75 & 0.75 & 0, 21, 40 & 0, 21 & 90, 88.1\\
NIRCam/F150W2 + F322W2 & 4.5-9.0 & 3500 & 5.0 & 2600-3200 (300) & 0.25 & 0.47 & 0, 40 & 0 & 90, 88.8\\
\hline
\end{tabular}
\end{center}
\end{table*}

We tested different choices for the spectral resolution, and reduced it until the SNR on the measured ``flux \text{bump}'' (increasing with decreasing spectral resolution) allowed us to achieve a few hundred kelvins precision on the retrieved starspot temperatures in the brightest star cases (as described later in Section \ref{res}). In this way, we chose $R \simeq 10$ for our spectra, both for NIRSpec and NIRCam. We modelled the transmission spectrum uncertainties of a $0.75 \, R_\mathrm{J}$ planet in front of a 5000~K, $0.75 \, R_\odot$ star and one of a $0.25 \, R_\mathrm{J}$ planet in front of a 3500~K, $0.47 \, R_\odot$ star. The details of the two scenarios are summarised in Table \ref{tabinput}. 

For each point of the transmission spectrum, the simulated transit depth uncertainty was used to derive the relative flux uncertainties in the corresponding transit. They were calculated as \citep[e.g.][]{sarkar2020}
\begin{equation}
    \sigma_s (\lambda) \simeq \frac{\sqrt{N_\mathrm{int}}}{2} \sigma_D(\lambda),
    \label{noise}
\end{equation}
where $N_\mathrm{int}$ is the number of exposures in the transit and $\sigma_D(\lambda)$ is the uncertainty on the transit depth at the same wavelength, as shown in Figure \ref{spec_unc}. We set an exposure time of 60 s for the simulated light curves, which given the transit parameters (described later in this Section) resulted in $N_\mathrm{int} = 432$ and 188 for the K and the M star simulation, respectively. We assumed our mock observations to be already corrected for systematic and astrophysical noise, and did not include the contribution from stellar granulation.

White noise in the wavelength-dependent transits results in the placement of each data point on a transit model to which a Gaussian scatter calculated with Equation \ref{noise} is added. By iterating our simulations, we noticed that different noise realisations have an impact on the final measure of $T_\bullet$. The implications of being limited, in an observational context, to a specific noise instance can be appreciated by running a large number ($\gtrsim 10$) of simulations, each one with a different noise realisation, and by merging the posterior distributions on the fitted parameters. Following \cite{feng2018}, we decided to estimate this effect by avoiding the inclusion of Gaussian scatter in the transit data points, and by later evaluating the impact of this choice on a few significant scenarios (Section \ref{noisediscussion}).
    
We then used \textsc{KSint} \citep{montalto2014} to simulate the transits affected by the occultation of an active feature, including the transit depth contamination due to the out-of-transit spot-induced stellar flux variation. This software takes as input the stellar density, which we calculated from the stellar radius and respective $\log g$ for each scenario (as presented in Table \ref{tabinput}). Given the focus on simulations of a single transit event, the planetary orbital period and stellar rotation period did not significantly impact our analysis: we then fixed the orbital period to 2 days and the stellar rotation period to 11 days (this latter choice represents an acceptable value for moderately active stars of the type we explored, \citealp[e.g.][]{mcquillan2014}). For all targets, we set no spin-orbit misalignment.

There exists a multitude of possible configurations that would produce a measurable starspot crossing, but here we have selected reasonable physically motivated scenarios that can highlight the potential of \textit{JWST} to provide further insights into starspot properties from planetary transits. To start with, we used a $3^\circ$-wide spot for our simulations, a value that sits between the one of large spot groups on the Sun \citep{mandal2017} and the largest spots observed for M dwarfs \citep{berdyugina2011}, and which is also consistent with starspots observed on active K stars \citep[e.g.][]{morris2017}. Secondarily, we chose several positions for the starspot: one close to the centre of the stellar disc ($\theta = 0^\circ, \, \mu = 1$) and one closer to the limb ($\theta = 40^\circ, \, \mu = 0.77$) at $0^\circ$ stellar latitude, associated to an $i=90^\circ$ planet orbital inclination. We also simulated spots at $21^\circ$ stellar latitude, corresponding to  $\theta=21^\circ$ ($\mu=0.93$), occulted by a planet at $i \simeq 88^\circ$. In this case, we introduced a slight displacement between the starspot and the projection of the planet centre on the stellar disc. In the hypothesis of circular starspots, the planet and starspot centres are aligned if
\begin{equation}
    i = \arccos \bigg(\frac{R_\star}{a} \sin \xi \bigg),
\end{equation}
where $a$ is the orbital semi-major axis and $\xi$ is the starspot latitude.

We used no starspot other than the occulted one, and the other transit parameters as in Table \ref{tabinput}. The orbits were assumed to be circular. 

\textsc{KSint} uses a quadratic LD law: we calculated the corresponding coefficients by fitting PHOENIX specific intensity spectra. For the NIRCam modes, we downloaded the throughput curves from the SVO Filter Profile Service\footnote{Described at https://jwst-docs.stsci.edu/near-infrared-camera/nircam-instrumentation/nircam-filters.} \citep{rieke2005,rodrigo2012,rodrigo2020}; those for NIRSpec (CLEAR filter) are available with the \textsc{Pandeia} reference data \citep{pontoppidan2016}. 

For each wavelength bin, we computed the starspot contrast as a function of wavelength as the ratio of two PHOENIX specific intensity models at different temperatures ($T_\bullet$ vs. $T_\star$) and using a starspot $\log g$ as low as the stellar $\log g - 0.5$ dex.

\begin{table}
\caption{Boundaries on the fitted parameters. $\mathcal{U}[a, b]$ and $\mathcal{J}[a, b]$ represent a Uniform and a Jeffreys prior between $a$ and $b$ (inclusive), respectively. $\mathcal{G}(a, b)$ denotes a Gaussian prior with mean $a$ and standard deviation $b$. Two different occultation mid-times were adopted according to the limb-angle $\theta$ of each scenario.}
\label{tabpriors}
\begin{center}
\begin{tabular}{lc}
\hline \hline
Parameter &  Prior \\
\hline
$R_\mathrm{p}/R_\star$ &  $\mathcal{J}[0.01, 0.2]$ \\
Orbit inclination $i$ [deg] & $\mathcal{U}[80, 90]$ \\
Transit mid-time $t_\mathrm{tr}$ [days] &  $\mathcal{U}[0.08, 0.12]$\\
LD linear coefficient $u_1$ & $\mathcal{G}(\mathrm{theoretical \,\,  value}, 0.05)$\\
LD quadratic coefficient $u_2$ & $\mathcal{G}(\mathrm{theoretical \,\, value}, 0.05)$\\
Gaussian peak $\Delta f$ & $\mathcal{U}[10^{-6}, 0.1]$ \\
Gaussian flattening factor $n$ & $\mathcal{U}[2, 8]$ \\
Gaussian width $w$ [days]& $\mathcal{U}[7 \times 10^{-4}, 0.02]$\\
Occultation mid-time $t_\bullet$ [days] ($\theta=0^\circ$) & $\mathcal{U}[0.09, 0.11]$\\
Occultation mid-time $t_\bullet$ [days] ($\theta=21$ or $40^\circ$) & $\mathcal{U}[0.11, 0.13]$\\
Out-of-transit quadratic term $r_0$ & $\mathcal{U}[-1, 1]$\\
Out-of-transit linear term $r_1$ & $\mathcal{U}[-1, 1]$ \\
Out-of transit level $r_2$ & $\mathcal{U}[0, 10]$ \\
\hline
\end{tabular}
\end{center}
\end{table}

\subsection{Transit fits}
We treated the synthetic data the same standard way that transits are typically analysed. We chose a transit modelling approach that allows for efficient iterative-based analyses that is insensitive to the degeneracies between starspot size, location on stellar disc, and inclination of planetary orbit. In place of using \textsc{KSint} for the transit fit we adopted the \textsc{batman} code, which implements the standard \cite{mandelagol2002}'s transit model \citep{kreidberg2015_batman}. We then expanded \cite{fraine2014}'s approach by adding the starspot occultation as a flattened Gaussian function profile to the transit. As a function of time $t$, this is equal to 
\begin{equation}
\Delta f \exp \bigg[ \bigg(- \frac{|t-t_\bullet|}{w} \bigg)^n \bigg],
\label{exp}
\end{equation}
where $\Delta f$ is the peak of the flux bump during occultation (the main parameter we were interested in), $t_\bullet$ is the occultation mid-time, $w$ is the width of the bump, and $n$ determines the flattening of the bump for starspots which are smaller than the projection of the planet on the stellar disc.

We fitted for the planet-to-star radius ratio $R_\mathrm{p}(\lambda)/R_\star$, LD coefficients, starspot bump peak $\Delta f$, and for a quadratic trend with time, $r_0 t^2 + r_1 t + r_2$, to account for the out-of-transit stellar flux modulation due to the presence of the starspot. As the orbital inclination $i$, the transit mid-time $t_\mathrm{tr}$ and the spot crossing mid-time $t_\bullet$ are common among all transits of the same transmission spectrum, we fitted these parameters on a band-integrated version (commonly referred to as ``white light curve'') of the transits for each scenario. We also assumed the bump width $w$ and flatness $n$ of the occultation bump to be independent of wavelength, and fitted for them only in the co-added transit. For the fit of the spectrophotometric transits in each scenario, we then fixed the value of these parameters to the median value obtained on the corresponding band-integrated transits.

Given the complexity of the transit and starspot parameter space, we sampled the parameter posterior distributions with nested sampling, which is more efficient than Markov chain Monte Carlo algorithms in dealing with multimodal posterior distributions \citep{skilling2006,higson2019}. We adopted the ``standard nested sampling'' implementation in the \textsc{dynesty} software \citep{speagle2020}, using Uniform priors for all transit parameters except $R_\mathrm{p}/R_\star$, for which we used a Jeffreys prior, and the LD coefficients, for which we used Gaussian priors centred on the theoretical values from stellar models. We imposed boundaries on the fitted parameters as indicated in Table \ref{tabpriors}; in particular, we used two different priors on $t_\bullet$ for the $\theta=0^\circ$ and the $\theta=40$ and $21^\circ$ scenario, to ease convergence. We started the sampling of the band-integrated transits with 200 live points, and reduced this number to 150 for the spectrophotometric transits.

The choice of fitting for the LD coefficients was due to the full transit coverage of future \textit{JWST} observations. Indeed, inaccurate and incomplete opacities in stellar atmosphere models, as well as small variations in the stellar atmospheric parameters due to brightness inhomogeneities on the stellar photosphere, make the LD coefficients derived from stellar models unreliable for active stars \citep[e.g.][]{csizmadia2013,maxted2018,heller2019}. Despite being small variations, such transit profile differences can have a measurable effect on the observed transit depths, and should be taken into account whenever possible. This said, we found that imposing a prior centred on the theoretical values of the LD coefficients provides better results than leaving wide Uniform priors. In our fits, we used the quadratic law in order to improve convergence with respect to the use of the four-coefficients law \citep{morello2017}.

\begin{figure}
\includegraphics[width=\columnwidth]{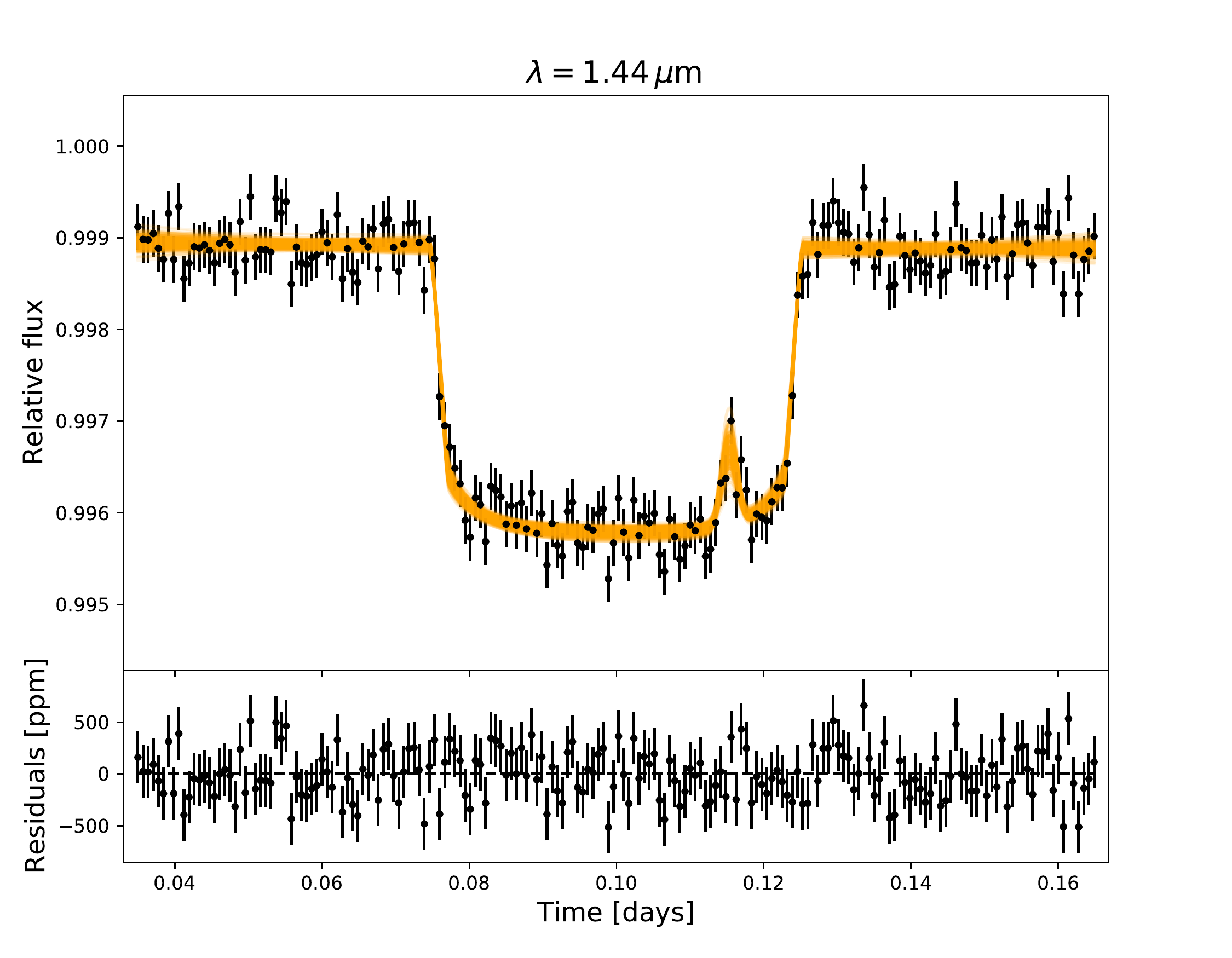}
\caption{Simulated transit of a 0.25 \rj~planet orbiting a 10.5-mag$_K$, 0.47 \rs, 3500~K star, occulting a 2900~K starspot at limb angle $\theta = 40^\circ$, observed with NIRSpec/Prism in the wavelength bin centred at $\simeq 1.4 \, \mu$m. This plot presents one of the noise realisations to better show the extent of the flux uncertainties. In orange on the upper panel, 300 samples from the posterior distributions obtained with the transit and Gaussian profile modelling are shown. The lower panel shows the residuals of the best-fit model, presenting a white-noise scatter, in parts per million.}
\label{transit_fit}
\end{figure}

Figure \ref{transit_fit} presents an example of fit on a simulated transit using the model just described; we there present one of the white noise realisations (Section \ref{noisediscussion}) to show an example of simulated transit uncertainties. The simultaneous transit and starspot fit was repeated for all wavelength bins and simulated stellar magnitudes, and the flux bump $\Delta f$ due to the starspot recorded.  

\subsection{Starspot contrast spectra fits} \label{spectrafit}
For each scenario, specific intensity spectra were used to fit the occultation bump peaks using Equation \ref{deltaf_f}. We obtained $F_\star$ using Equation \ref{fluxfromi}, relied on the measured $(R_\mathrm{p} (\lambda)/R_\star)^2$ and used the specific intensity models $I_\star(\mu_\bullet)$ and $I_\bullet(\mu_\bullet)$ at the same $\mu_\bullet$ chosen in input, assuming that $\mu_\bullet$ can be accurately determined from the observed $t_\bullet$.

For the stars, we selected a spectrum with the same parameters used in input. For the starspots, we interpolated the stellar models in the range 2300-3400 K for the M star ($T_\star = 3500$~K), and in the 3600-4900 K range for the K star ($T_\star = 5000$~K), using spectra with the same [Fe/H] as the input stellar spectrum and the stellar $\log g - 0.5$ (see Section \ref{starspotspectrum}). To do this, we performed a bivariate spline approximation with \textsc{scipy}'s  \textsc{RectBivariateSpline} implementation \citep{2020SciPy-NMeth}.

\begin{figure}
\includegraphics[width=\columnwidth]{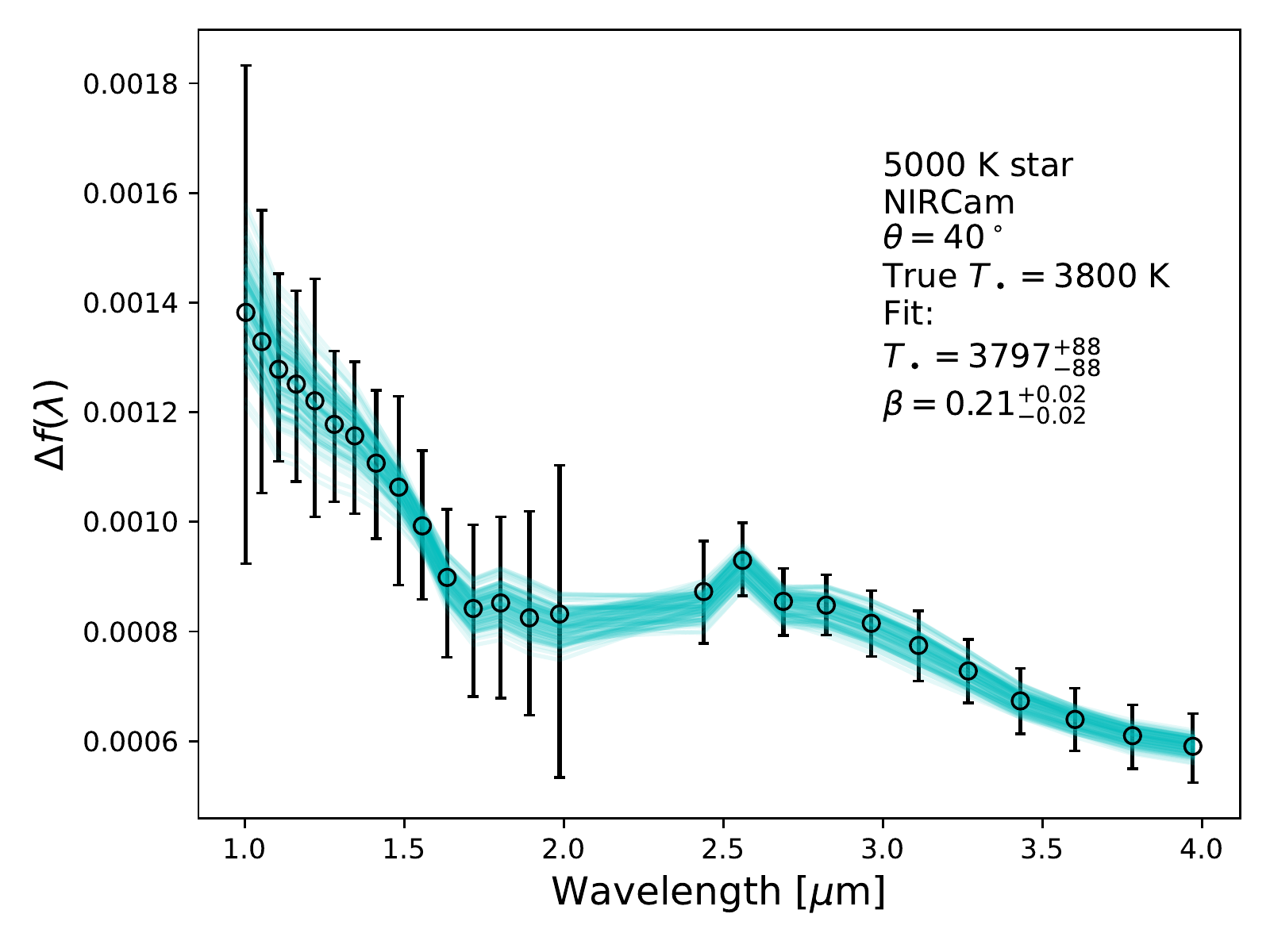}
\caption{Example of in-transit starspot crossing flux bumps measured with the model in Equation \ref{deltaf_f} and their uncertainties, for the mag $K=7.5$, K star case, in the $T_\bullet=3800$~K, $\mu$-angle $\theta=40^\circ$ scenario without in-transit Gaussian scatter. The models in light blue represent 100 fitting models sampled from the $T_\bullet$ and $\beta$ posterior distributions. The result of the nested sampling for fitted parameters, and their 68\% credible intervals, are indicated in the label. The model spectra were convolved with the NIRCam filters and degraded to the same resolutions as the observations before performing the fit.}
\label{spotspectrum}
\end{figure}

It is well known that the starspot effective temperature and filling factor (here probed by $\beta$) are strongly correlated parameters, as a cool and small starspot can cause the same in-transit bump peak as a large and warm spot. In particular, in some of the low-contrast cases we observed bimodal posterior distributions for these two parameters (see Section \ref{res_deg}). However, observations provide us with a constraint on the lower limit of $\beta$, and sometimes allow the choice of one of the modes. To find this value $\beta_\mathrm{min}$, we first derived the half-width at half maximum (HWHM$_\bullet$, in time units) of the Gaussian function which we used to describe the occultation bumps in the white light transits. Given Equation \ref{exp}, this was obtained as
\begin{equation}
    \mathrm{HWHM}_\bullet = w(\ln{2})^{1/n}.
    \label{hwhm}
\end{equation}
Then, assuming a circular starspot and zero-eccentricity orbit, we determined the minimum starspot radius $R_{\bullet, \mathrm{min}}$ consistent with the observed occultation, by inverting
\begin{equation}
    R_\mathrm{p} + R_{\bullet, \mathrm{min}} \simeq \frac{3}{2} \mathrm{HWHM}_\bullet v_\mathrm{p} = \frac{3}{2} \mathrm{HWHM}_\bullet \frac{2 \pi a}{P},
\label{starspotsize}
\end{equation}
where $v_\mathrm{p}$ is the tangential velocity of the planet in its orbit, $P$ is the planet's orbital period, and we approximated the last moment when the profiles of planet and starspot are in contact with $t_\bullet + 3 \,  \mathrm{HWHM}_\bullet /2$. By dividing both terms in Equation \ref{starspotsize} by the stellar radius and the measured planet-to-star radius ratio and taking their square, after some algebra we obtained an expression which is only based on measured quantities:
\begin{equation}
    \beta_\mathrm{min} \equiv \bigg( \frac{R_\mathrm{\bullet, \mathrm{min}}}{R_\mathrm{p}} \bigg)^2= \bigg[ \frac{3 \pi a/{R_\star}}{P (R_\mathrm{p}/R_\star)} \mathrm{HWHM}_\bullet - 1 \bigg]^2,
\label{starspotsize2}
\end{equation}
where $a$ is the orbital semi-major axis. This value is underestimated, as the measured transit depth is affected by the presence of starspots, but we consider it to be a sufficient approximation.

We sampled the $T_\bullet$ and $\beta$ posterior distributions with nested sampling, using Uniform priors for both parameters and starting the sampling with 100 live points. The starspot $T_\bullet$ was let vary across the full interpolated range (different for the M and the K star case), and $\beta$ was let float between $\beta_\mathrm{min}$ and 1 (inclusive). Figure \ref{spotspectrum} presents 100 fitting models from the posterior distributions for one of the K star scenarios without Gaussian scatter, simulated for NIRCam and downgraded to the resolution of the observations.

\begin{figure*}
\includegraphics[width=0.8\columnwidth]{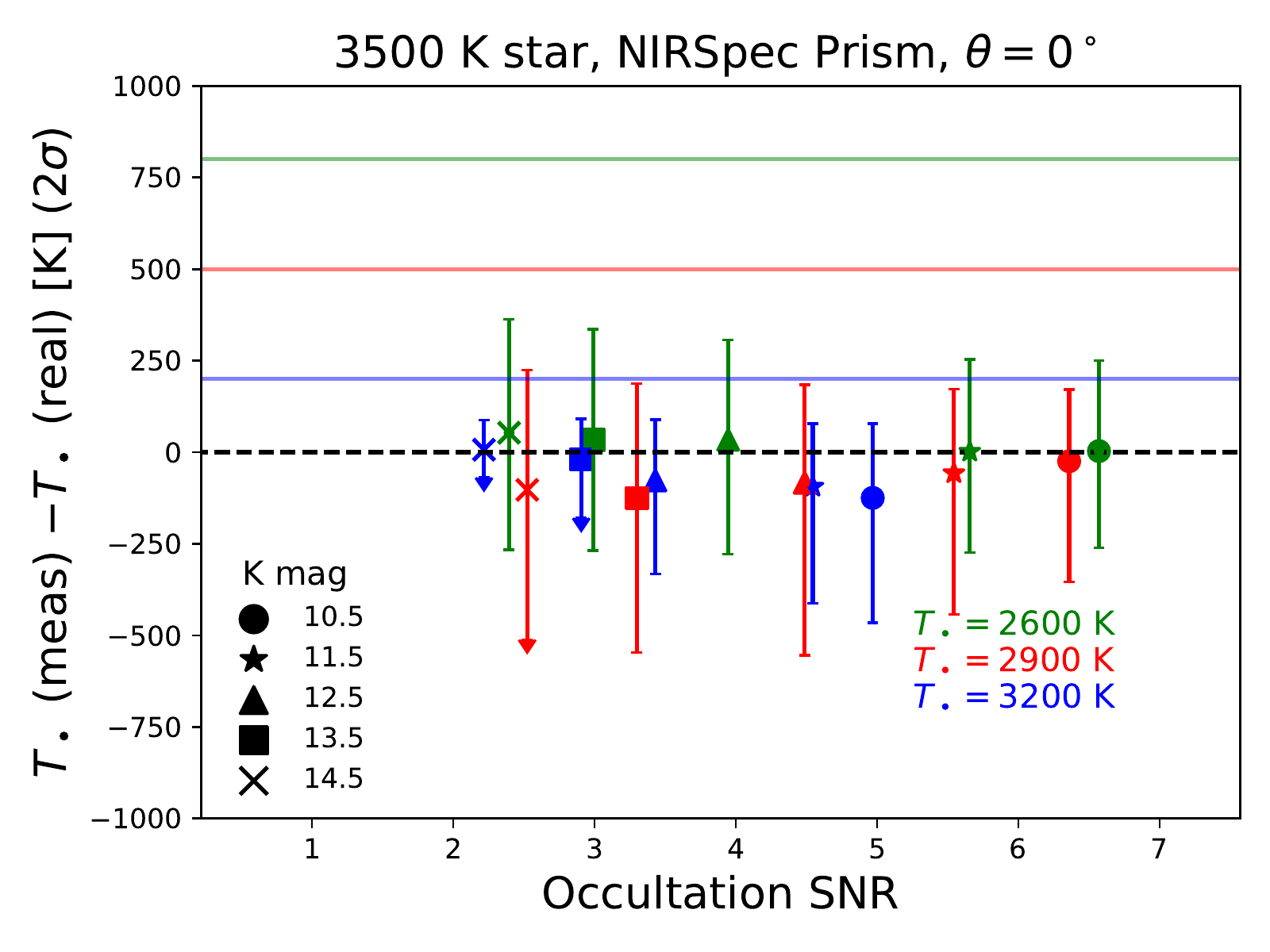}
\includegraphics[width=0.8\columnwidth]{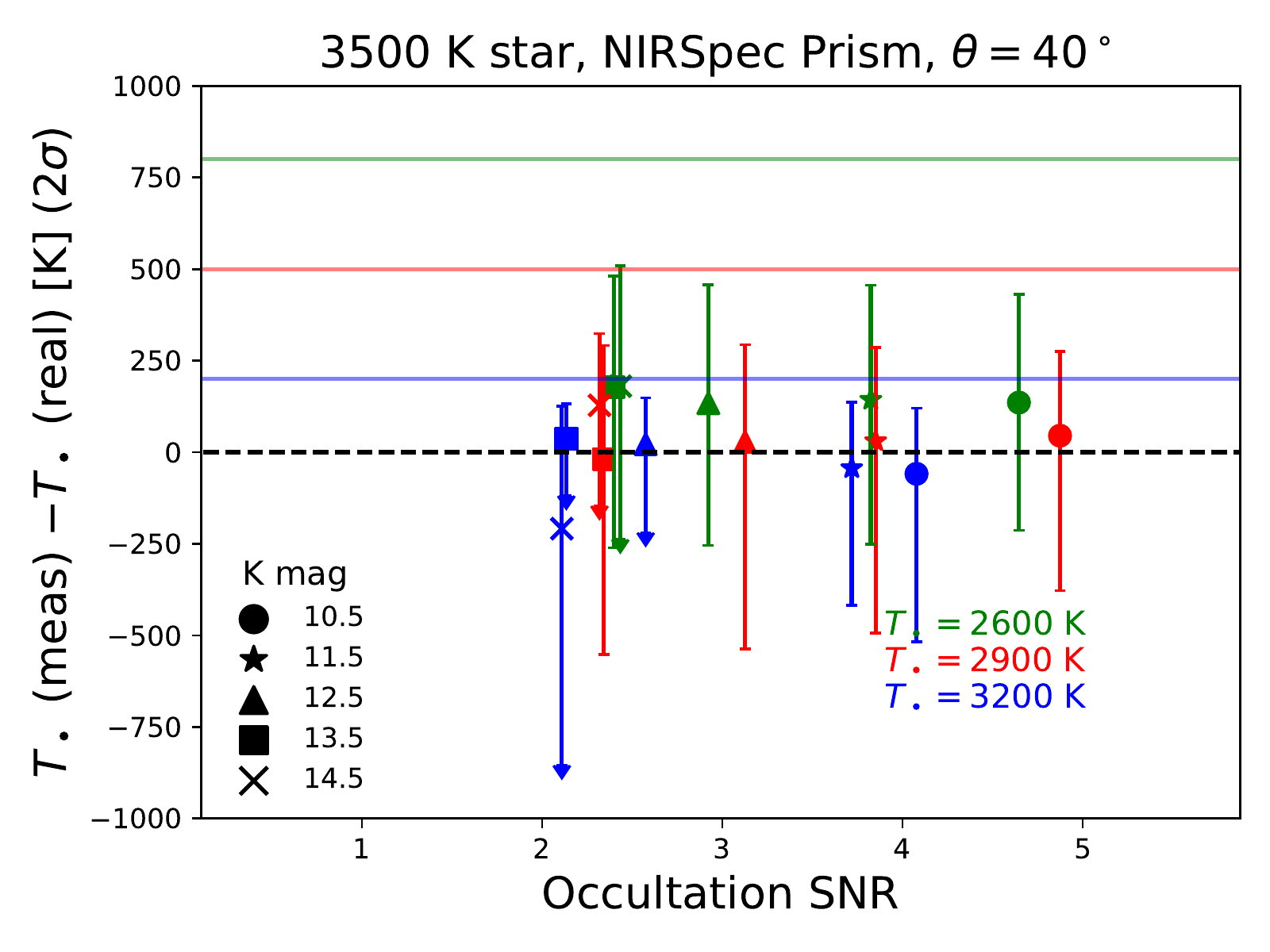}
\includegraphics[width=0.8\columnwidth]{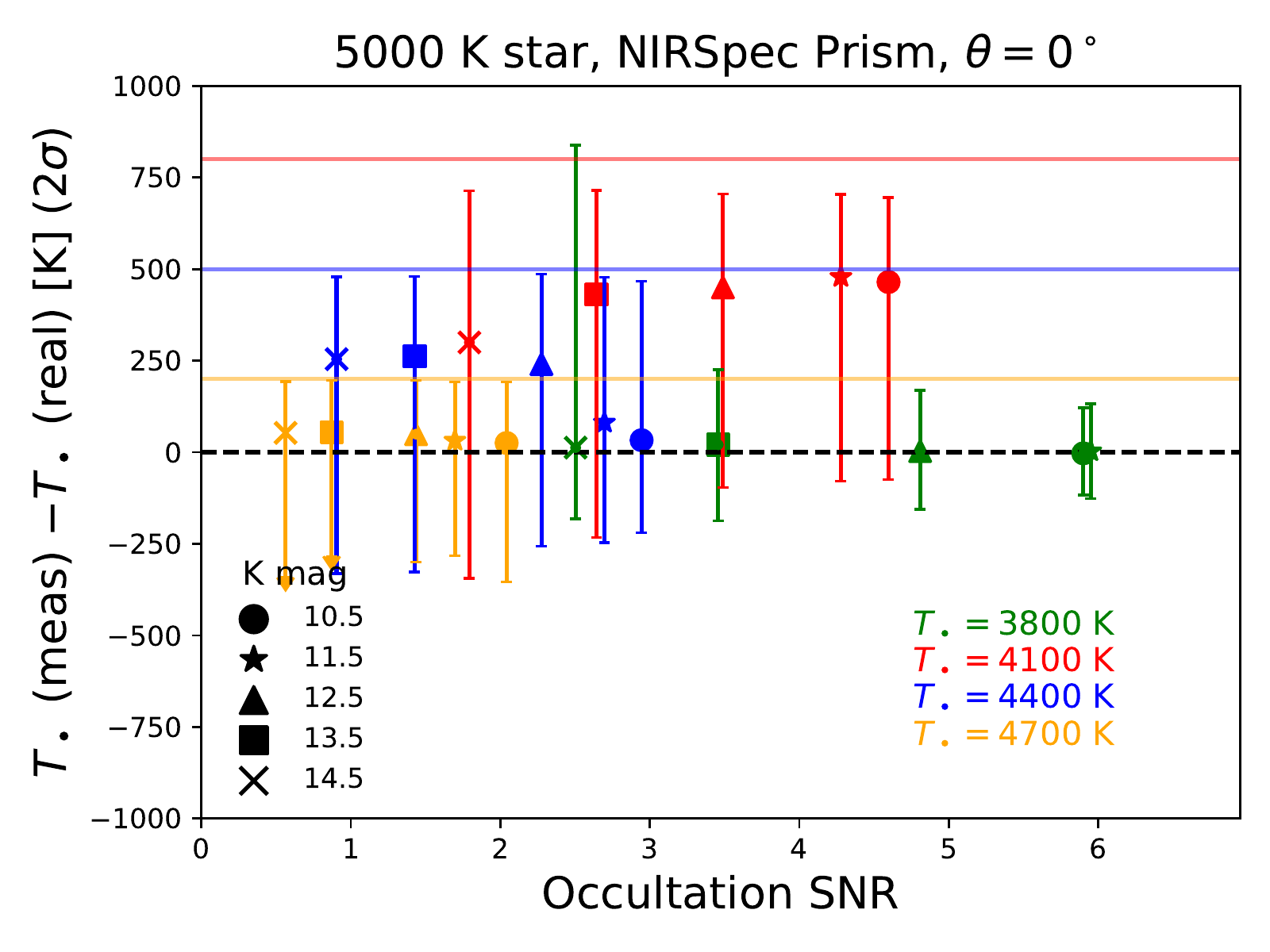}
\includegraphics[width=0.8\columnwidth]{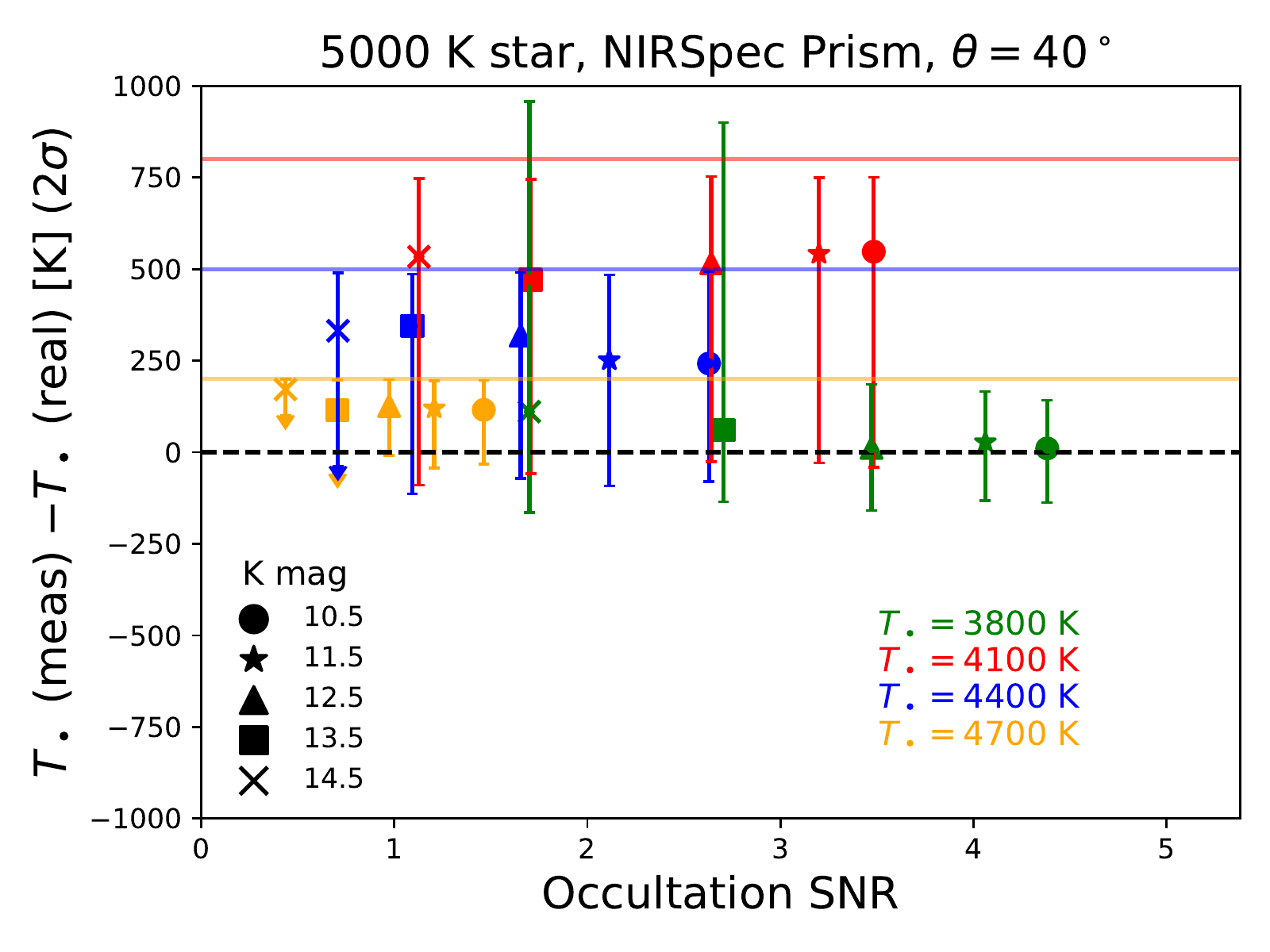}
\caption{Difference between the retrieved starspot temperature $T_\bullet \, (\mathrm{meas})$ and the true input $T_\bullet \, (\mathrm{true})$ ($y$-axis) compared to the occultation SNR ($x$-axis), for the scenarios simulated with NIRSpec/Prism using starspots at $0^\circ$ latitude; 95\% percentiles are reported. The stellar $K$ magnitudes corresponding to the different cases and the true $T_\bullet$ values are indicated with different markers and colours, respectively. The minimum allowed $T_\bullet$ in our retrievals, $T_\star - 100$~K, is indicated by a horizontal line with the same colour as the data points for a given scenario (the value for the $T_\bullet = 3800$~K falls off the plot edges). In the title of each plot, the simulated instrument, $T_\star$, and the limb angle $\theta$ are reported. Markers with arrows pointing down -- such as the lowest-SNR point in the top-left panel -- represent cases where the occultation flatness parameter $n$ is poorly determined ($\sigma_n \geq 2$), so that the error bars on $T_\bullet$ cannot be considered reliable (see Section \ref{lowsnr}).}
\label{resnirspec}
\end{figure*}

\begin{figure*}
\includegraphics[width=0.8\columnwidth]{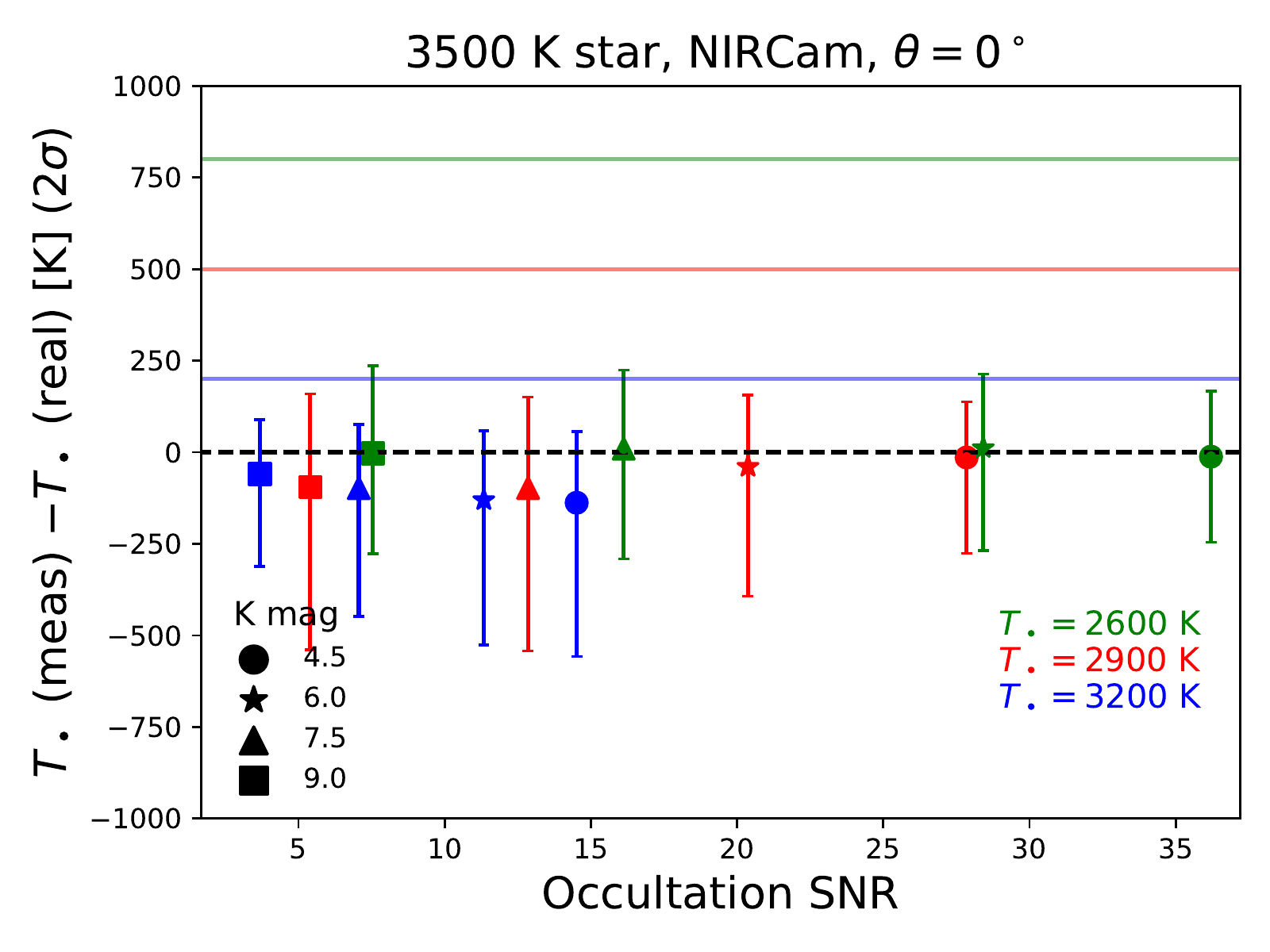}
\includegraphics[width=0.8\columnwidth]{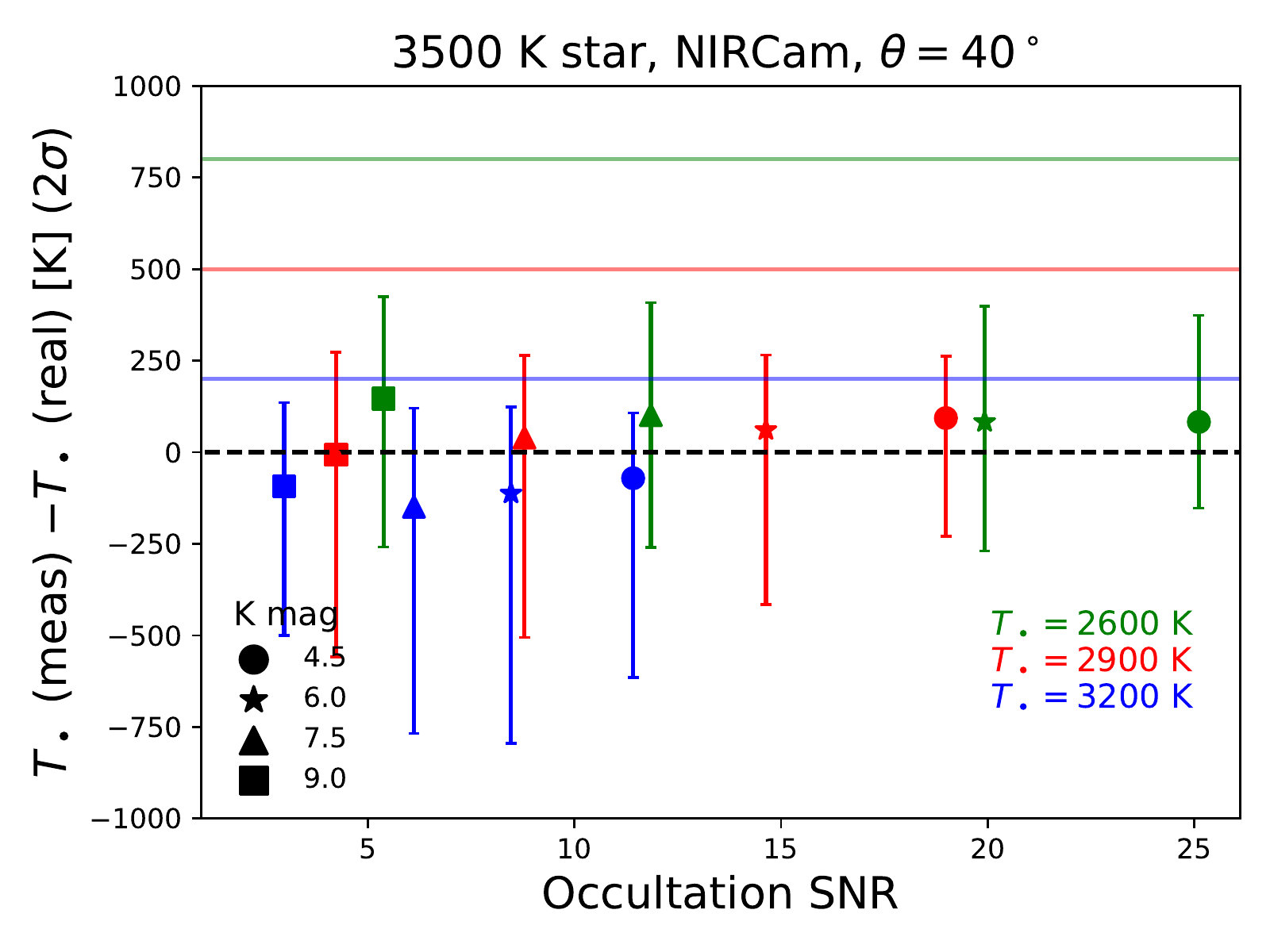}
\includegraphics[width=0.8\columnwidth]{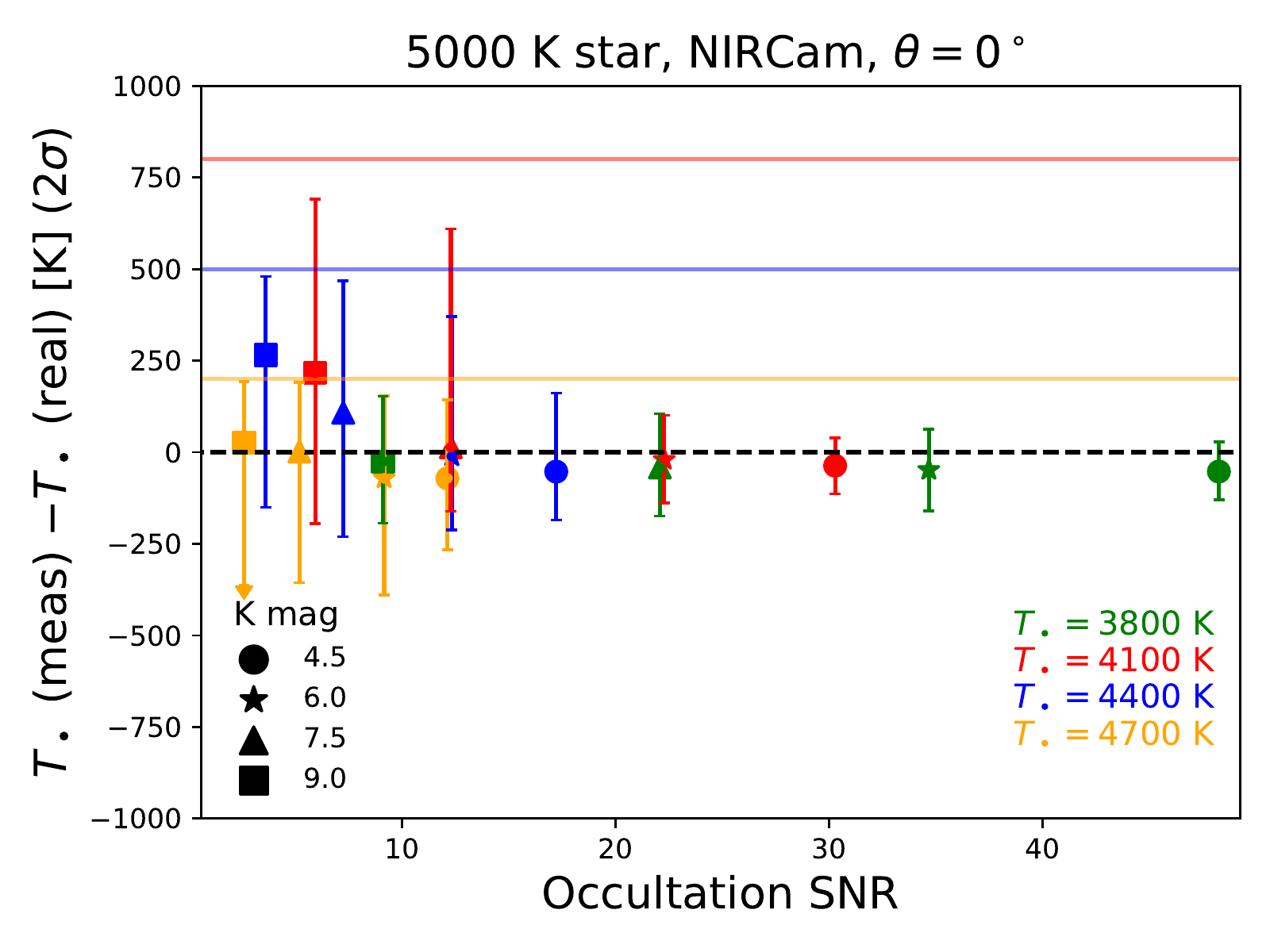}
\includegraphics[width=0.8\columnwidth]{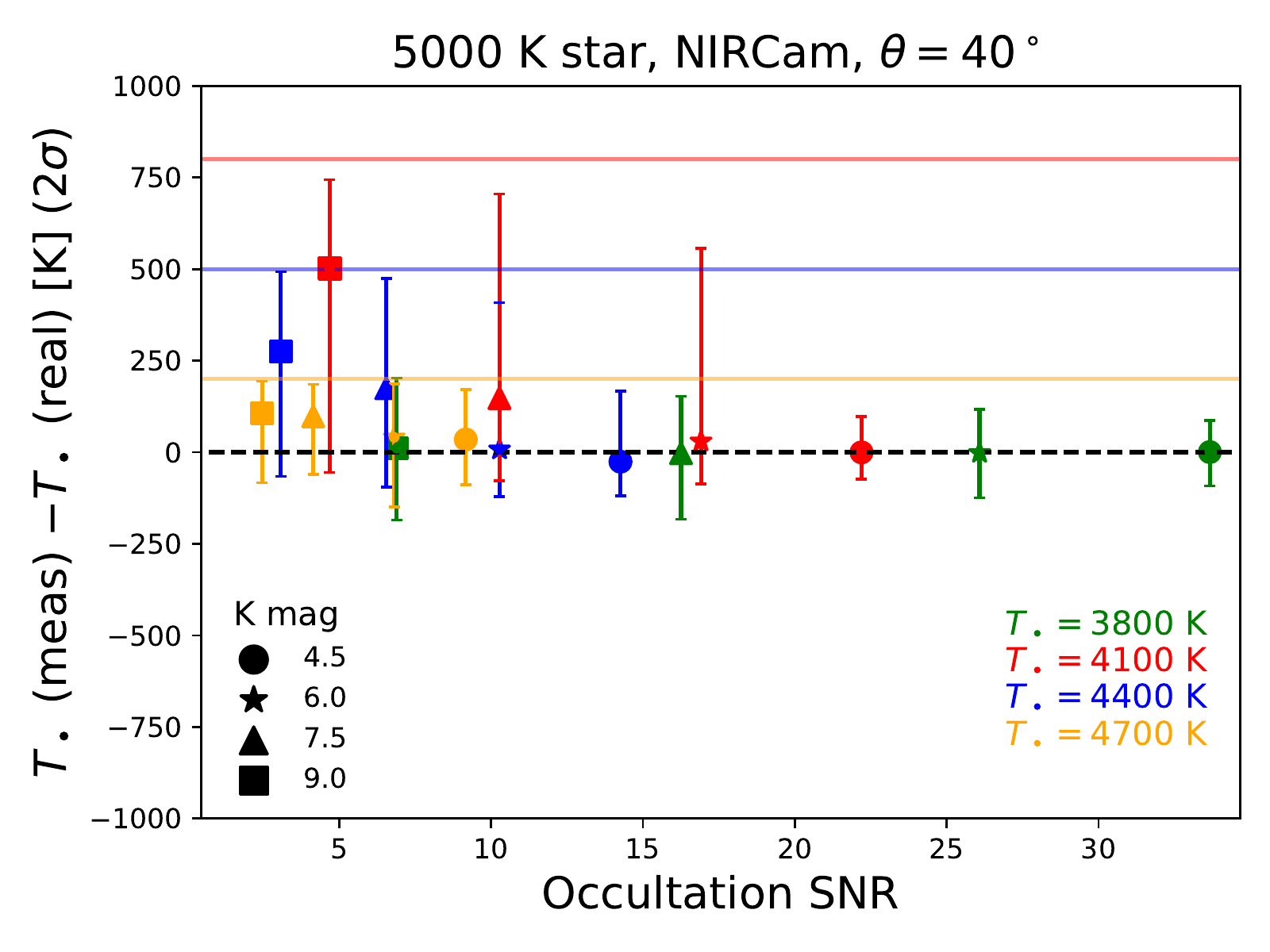}
\caption{Same as Figure \ref{resnirspec}, but for NIRCam/F150W2 + F322W2.}
\label{resnircam}
\end{figure*}

\begin{figure*}
\includegraphics[width=0.8\columnwidth]{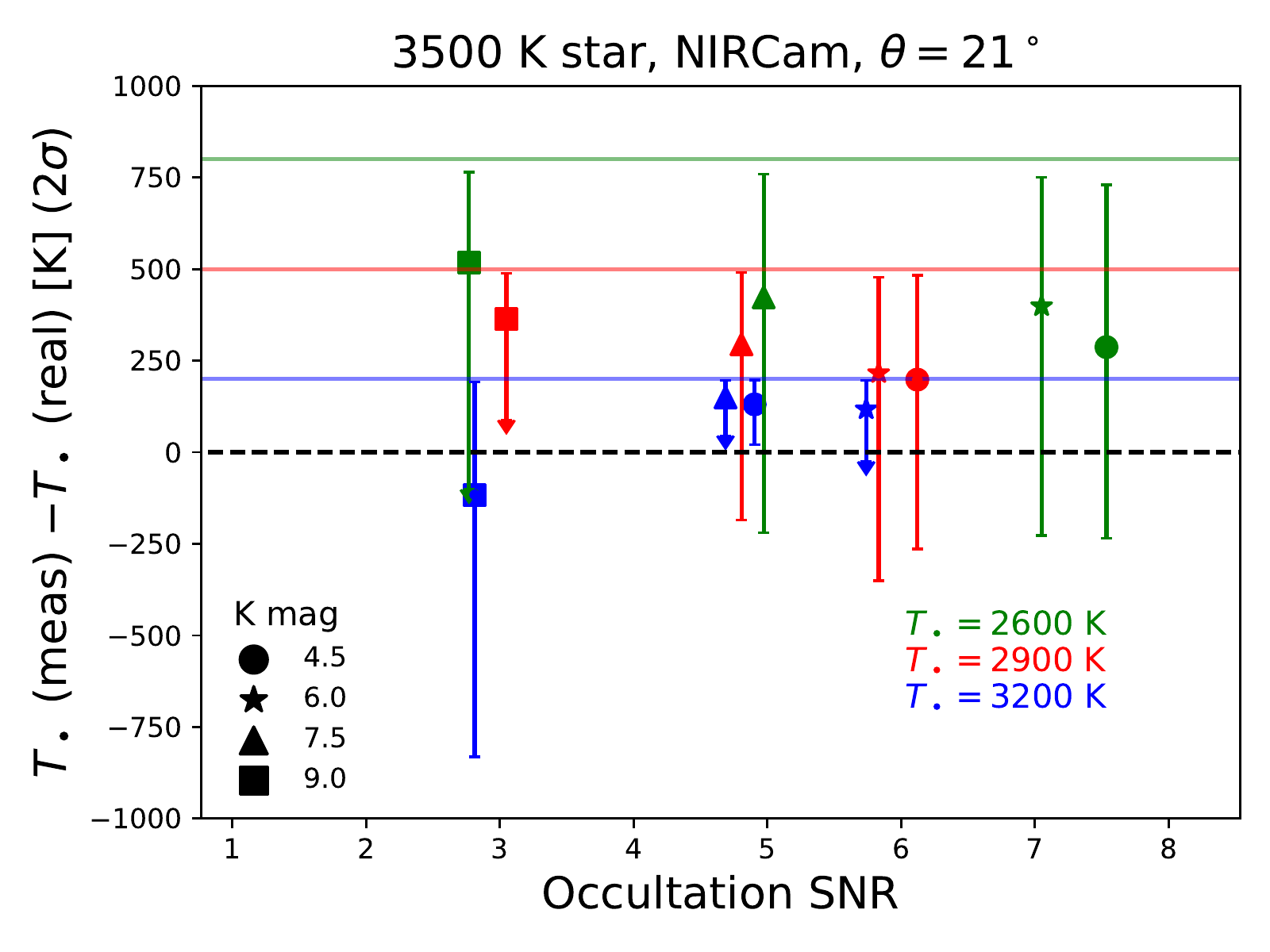}
\includegraphics[width=0.8\columnwidth]{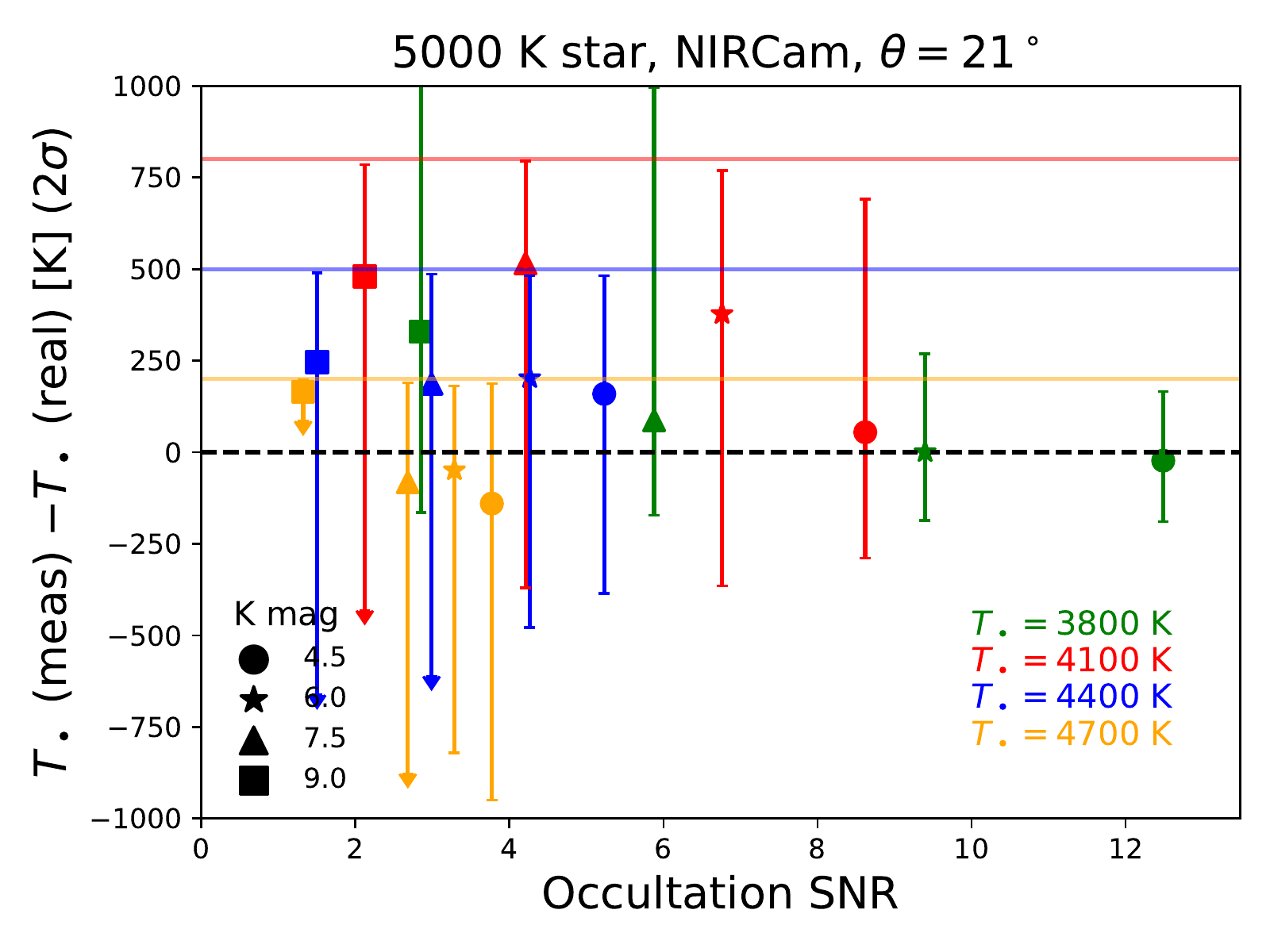}
\caption{Same as Figure \ref{resnirspec} for spots at $21^\circ$ stellar latitude, for the M (left) and K (right) scenarios observed by NIRCam/F150W2 + F322W2.}
\label{reslat}
\end{figure*}

\begin{figure*}
\includegraphics[width=\columnwidth]{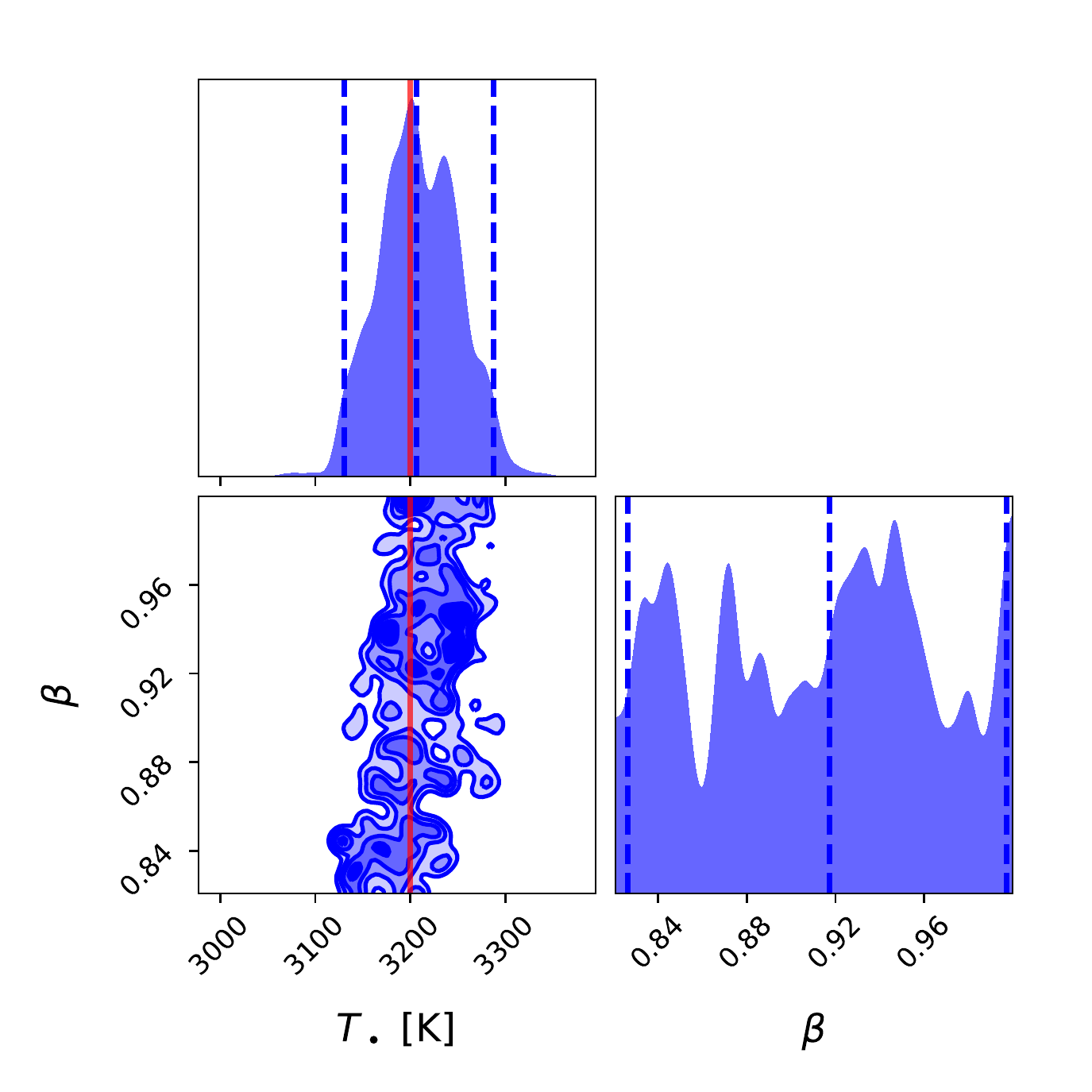}
\includegraphics[width=\columnwidth]{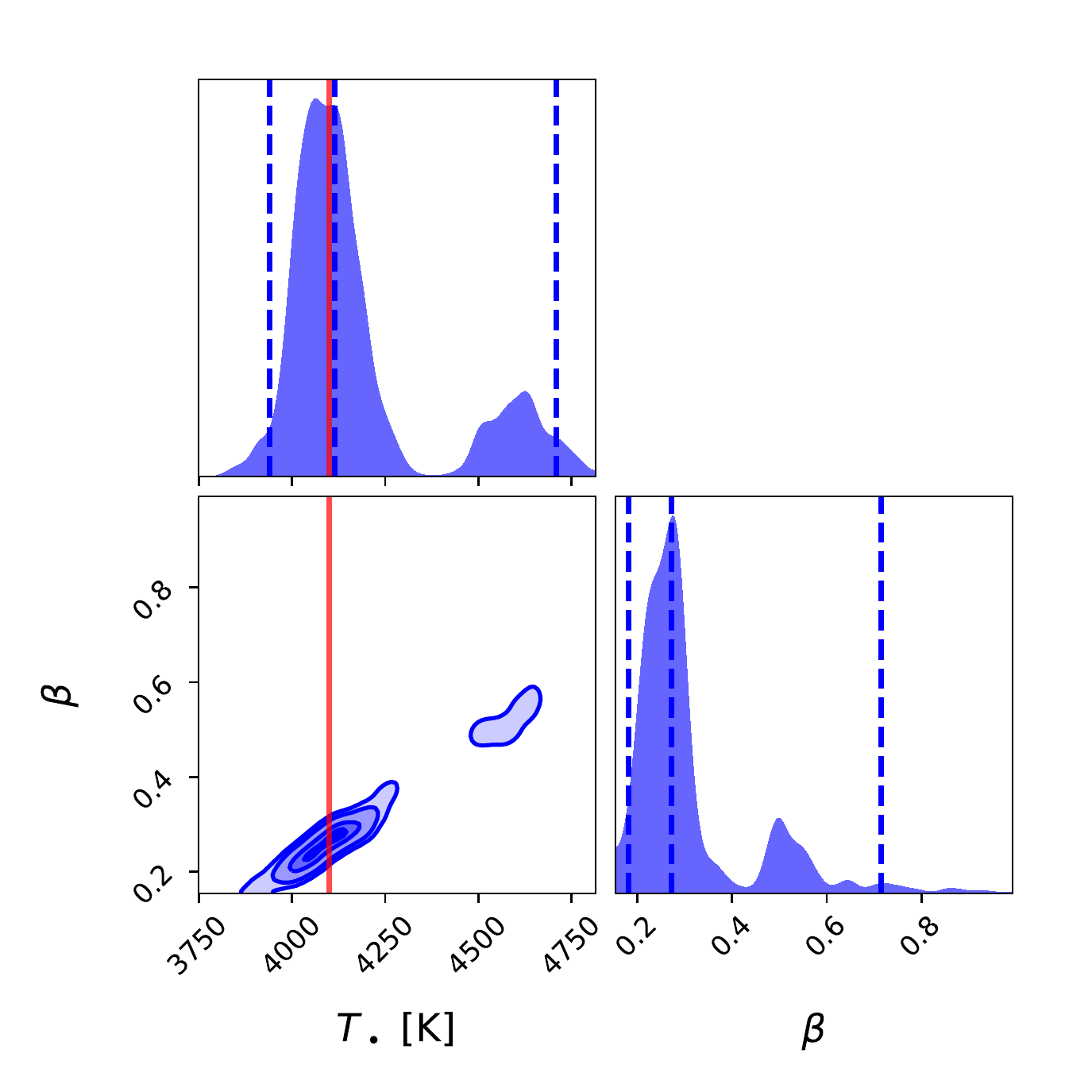}
\caption{\textit{Left:} $T_\bullet$ and $\beta$ posterior distributions for the 14.5 $K$ mag, M star observed with NIRSpec/Prism and a $T_\bullet = 3200$ K occulted starspot, with $\theta=0^\circ$. While the occultation signal is strong enough to exclude $T_\bullet = T_\star$, the high $\beta_\mathrm{min}$ forces the solution towards a warm spot (see Section \ref{lowsnr}). \textit{Right:} $T_\bullet$ and $\beta$ posterior distributions for the 7.5 $K$ mag, K star observed with NIRCam/F150W2 + F322W2 and a $T_\bullet = 4100$ K occulted starspot, with $\theta=0^\circ$ (see Section \ref{res_deg}). The $95\%$ credible intervals are represented with blue vertical dashed lines, while the true $T_\bullet$ value is indicated with a red vertical line.}
\label{posterior}
\end{figure*}

\section{Results}\label{res}
We present our results as a function of the SNR of the starspot spectrum derived from the transit fits. To do so, we defined a metric that contains information on the starspot contrast decrease from the optical to the NIR, on the contrast ``baseline'' in the white-light transit, and on the uncertainty in the $\Delta f(\lambda)$ values measured from the spectrophotometric transits. If the starspot contrast spectrum results from $N$ spectrophotometric transit light curves, corresponding to increasing wavelength in bins from 1 to $N$, the white-light transit is marked with $\left \langle N \right \rangle$, and $\Delta f_i$ is the bump peak measured in wavelength bin $i$, we computed
\begin{equation}
    \mathrm{SNR} = \frac{\sqrt{(\Delta f_2 - \Delta f_{N - 1})^2 + \Delta f_{\left \langle N \right \rangle}^2}}{\Delta (\Delta f_{N - 1})},
    \label{snr}
\end{equation}
where the term on the denominator is the uncertainty on the bump peak in the $(N - 1)$-th wavelength bin. We avoided the use of the shortest and the largest-wavelength transit, as these are the noisiest in the simulated transmission spectra (Figure \ref{spec_unc}).  

As the SNR of the bump signal increases for larger activity features, this representation adapts to scenarios with varying starspot sizes, even if we here discuss only the simulations we carried out with $3^\circ$-wide spots; in particular, the effect of a large spot on the occultation SNR is similar to the one of a bright star. 

\subsection{Starspots at $0^\circ$ latitude, planet with $i=90^\circ$}
The overall result of this analysis is that, despite their reduced contrast in the NIR compared to the visible, the effective temperature of starspots observed with \textit{JWST} can be constrained with an accuracy of a few hundred kelvins in the brightest star and largest contrast cases, and particularly with NIRCam. Figures \ref{resnirspec} and \ref{resnircam} present the retrieved values for the temperature contrasts and their 95\% credible intervals, as a function of the occultation (Equation \ref{snr}). The stellar photospheric values are also represented, and indicate when the derived $T_\bullet$ cannot be significantly distinguished from $T_\star$.

At $0^\circ$ latitude, the limb-angle position of the starspot ($\theta=0$ or $40^\circ$) slightly affects the scatter of the solution with respect to the correct results. For NIRCam/F150W2 + F322W2, simulations produced accurate results for the coolest starspot cases, even for dim ($K=9$) magnitudes; with NIRSpec/Prism, a few hundred kelvin precision was achieved for the largest temperature contrasts ($T_\star - T_\bullet = 900$ and $1200$~K for the M and the K star, respectively) for $K\leq12.5$ magnitudes. For both the M and the K star, the constraints on the temperature contrast become weaker as the true $T_\bullet$ approaches the photospheric temperature, even if the occultation bump is detected in the transits. For the lowest SNR cases, our simulations were only able to place a lower limit to the spot temperature; in the mid-SNR cases, the starspot temperature-size degeneracy caused the $T_\bullet$ posterior distribution to include both very close values to $T_\star$, as well as several hundred kelvin lower values (see Section \ref{res_deg}). We notice that $T_\star - T_\bullet = 1200$~K contrasts are realistic for K dwarfs, while larger contrasts than 600~K are unlikely on M stars \citep{berdyugina2005,herbst2021}.

\subsection{Starspots at latitude $> 0^\circ$, planet with $i \neq 90^\circ$}
As described in Section \ref{transitmodelling}, we investigated the impact of a different geometrical configuration, including a slight displacement between the starspot and the planet projected centres. For the NIRCam simulations, which include the largest SNR cases, we simulated a starspot at $21^\circ$ stellar latitude (requiring also $\theta=21^\circ$), and added a $\lesssim 1^\circ$ angle to the planet orbital inclination $i$ with respect to the value which results in perfect alignment. Figure \ref{reslat} shows that, as expected, the lower SNR of the occultation results in larger scatter and error bars for the retrieved $T_\bullet$. We found best results for the largest contrast and the brightest ($K=4.5$ and 6.0) cases, particularly for the K star.

\subsection{Low-SNR solutions}\label{lowsnr}
We observed that solutions with SNR$\lesssim 3$ are driven by the choice of priors and cannot be considered reliable. As an example, the solution for the $T_\bullet=3200$~K, $\theta=0^\circ$ spot on the mag $K=14.5$ M star observed with NIRSpec/Prism (Figure \ref{resnirspec}, top-left panel) has smaller error bars compared to the simulations with the same stellar and spot parameters but brighter stars. This happens because the bump flatness parameter $n$ (Equation \ref{exp}) is not constrained by the white-light transit fit. The median value of its posterior distribution, used to derive $\beta_\mathrm{min}$ through Equations \ref{hwhm} and \ref{starspotsize2}, is then found close to values around the middle value of the prior given to $n$ (set between 2 and 8, as shown in Table \ref{tabpriors}; Figure \ref{corner_transit} illustrates the posterior distributions for this transit fit). As a result, $\beta_\mathrm{min}$ is found to be close to 1 (i.e., the spot must be almost or completely occulted by the planet) and good fits to the spot contrast spectrum are only those produced by a ``warm'' activity feature, as shown in the left panel of Figure \ref{posterior}. Hence, even if the occultation signal is strong enough to exclude a $T_\bullet$ as high as $T_\star$ (otherwise, only a lower limit could be attributed to $T_\bullet$), the solution is ``forced'' around warm spots.

To confirm this, such a solution cannot be reproduced if the constraint on $\beta_\mathrm{min}$ is lifted, i.e. if a (0, 1] prior is set for $\beta$. We therefore consider this solution not reliable, and recommend the inspection of the occultation bump fit in white-light transits to correctly interpret the derived $T_\bullet$ values. Cases with a poor fit for the occultation flatness parameter $n$ (defined as where its error bar $\sigma_n$ is $\geq 2$) are marked using arrows pointing down in Figures \ref{resnirspec}, \ref{resnircam} and \ref{reslat}, to indicate that cooler spot solutions cannot be excluded.

\subsection{Starspot temperature-size degeneracy}\label{res_deg}
For warmer starspots and dimmer stars, the SNR of the occultation feature decreases, and the fitting procedure struggles in distinguishing between the starspot temperature and the stellar photosphere's. As the lowest allowed $T_\bullet$ in our model was $T_\star - 100$~K, we required this latter value to be excluded at the 95\% confidence level for the temperature contrast to be detected. The low-SNR cases without such detection were clearly affected by the starspot contrast and size degeneracy, contrarily to the large SNR cases. In some of these cases, the $T_\bullet$ posterior distribution spans a large part of the prior space and does not lead to significant constraints on this parameter. We observed that, in some specific scenarios, the $T_\bullet$ posterior distribution is bimodal, with the correct value lying in one of the modes. One example is the case of the $T_\bullet = 4100$~K starspot on the K star: here, the correct mode cannot be chosen by using the method presented in Section \ref{spectrafit}, as the $\beta$ values of both modes are enclosed in the prior. Figure \ref{posterior}, right panel, shows this situation for the mag$_K = 7.5$ star observed with NIRCam/F150W2 + F322W2; in a case like this, the 50\% percentile shown in Figures \ref{resnirspec}, \ref{resnircam} or \ref{reslat} is little representative of the shape of the posterior. In Appendix \ref{appendix_posteriors}, we provide the posterior distributions for all the simulations, for a more thorough examination of our results: in particular, we highlight that the bimodality just described is not found at the largest SNR values. 

We remark that some arguments could be used to choose between the modes of the posteriors. For example, the lowest mode in Figure \ref{posterior}, right panel, has $\beta \equiv (R_\mathrm{p}/R_\bullet)^2 = (R_\mathrm{p}/R_\star)^2(R_\star/R_\bullet)^2 \simeq 0.5$. Using the measured transit depth value ($\simeq 0.01$), this means that the angular size of the starspot for this solution is $\simeq 8^\circ$. Constraints on the largest likely value for the starspot size, which we do not examine here, could then be applied.

\subsection{Implications of a single noise realisation}\label{noisediscussion}
To explore the consequences of using transits without Gaussian scatter for our simulations (Section \ref{transitmodelling}), we selected a few scenarios resulting in significant constraints on $T_\bullet$, and repeated for ten times our simulations using each time a different noise realisation. Here we discuss the case of the mag $K=4.5$ K star with a 3800~K starspot and $\theta=0^\circ$, observed with NIRCam/F150W2 + F322W2. For each noise instance, we recorded the posterior distributions of $T_\bullet$ and $\beta$, and then compared the merged posterior distributions to those obtained without scatter in the transits.

\begin{figure*}
\includegraphics[width=\columnwidth, trim=0 2cm 0 0]{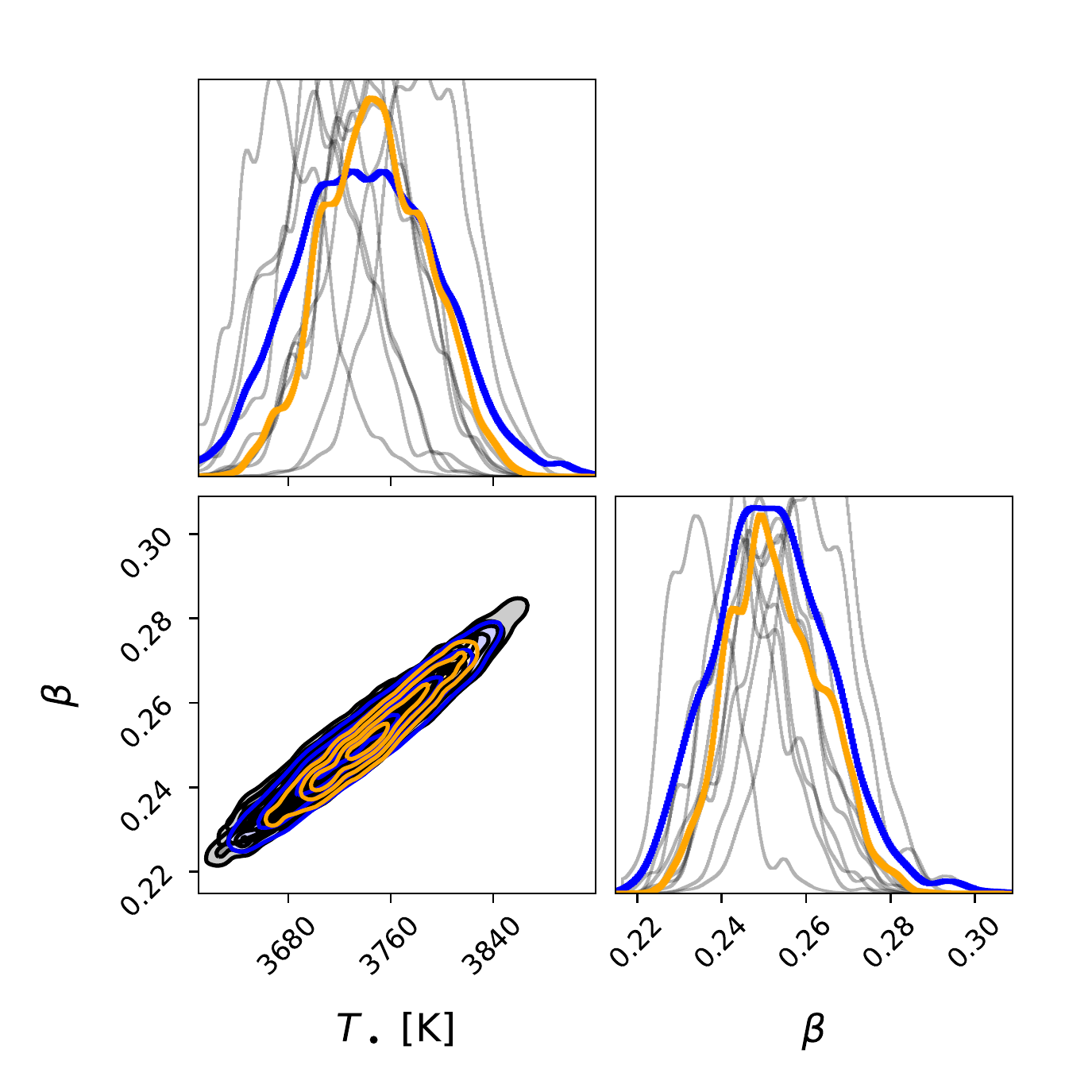}
\includegraphics[width=\columnwidth]{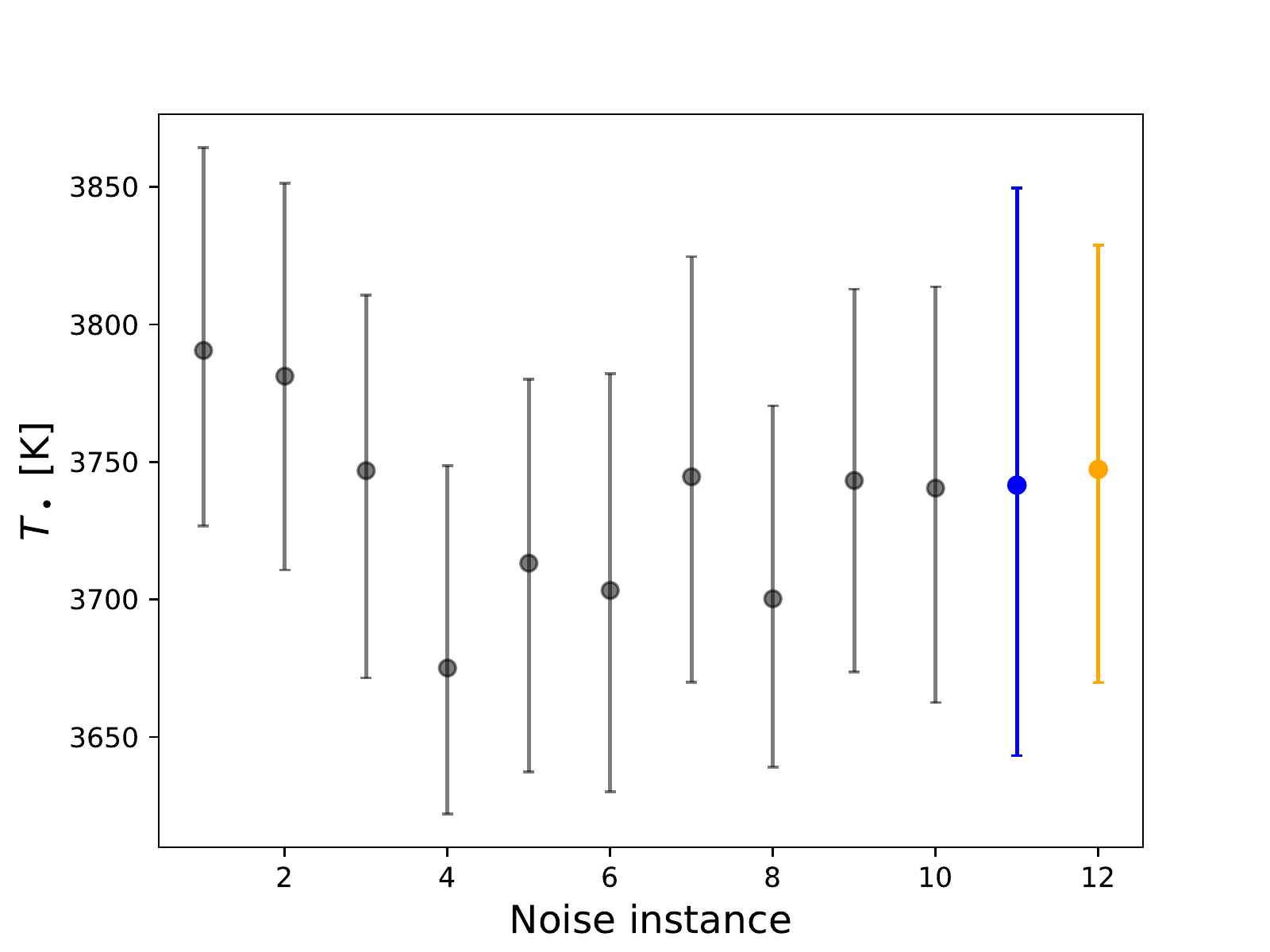}
\vspace{1cm}
\caption{$T_\bullet - \beta$ posterior distributions (\textit{left}) and $T_\bullet$ 95\% percentiles (\textit{right}) derived from the transits without Gaussian scatter and with ten noise realisations from one of our high-SNR simulations (see Section \ref{noisediscussion}). Individual noise instances are marked in light grey, their merged distribution in blue (noise instance 11 in the right panel), and the case without Gaussian scatter in orange (noise instance 12 in the right panel).}
\label{mergedhisto}
\end{figure*}

Figure \ref{mergedhisto} presents the posterior distributions derived with this method and the resulting 95\% percentiles. It can be noticed that, in most cases, the medians of the distributions are within each-other's uncertainties; the median values themselves vary by no more than 10\% from one instance to the other, as well as compared to the case without Gaussian scatter. However, the merged posterior distribution shows -- as expected -- an increase in the uncertainties, which indicates that error bars drawn from a single noise instance are too optimistic. In this specific case, we observed a 20-30~K increase in the largest and lowest $T_\bullet$ values allowed within the 95\% percentiles. Hence, while one might be confident on the median $T_\bullet$ values achieved with the method we presented, the uncertainties should be regarded as likely underestimated in the mid- to high-SNR cases.

\section{Discussion and conclusions}\label{disc}

Our simulations show that \textit{JWST} can be expected to be competitive for the determination of starspot temperatures from transit data with respect to \textit{HST}. This latter was able to provide $\lesssim 250$~K uncertainties on starspots occulted by planets transiting very bright stars, such as the mag$_K \simeq 5.5$ HD 189733, by using observations in the visible \citep{sing2011}; about $400$~K precision was also shown to be achievable, by stitching together visible and near-infrared spectra affected by unocculted spots \citep{bruno2020,rathcke2021}. The broad wavelength coverage of \textit{JWST} will be able to place significant constraints by relying on near-infrared observations alone. Moreover, another advantage of \textit{JWST} will be the ability to observe full transits without interruptions, thanks to its positioning in orbit around L2: this will allow a better coverage of transit events, both at the transit centre and edges.

With our method, we achieved $\lesssim 250$ K uncertainties on K and M stars with cool starspots observed with NIRCam/F150W2 + F322W2, as well as on mag$_K \leq 12.5$ stars observed with NIRSpec/Prism. Large-contrast starspots highlight indeed important temperature-sensitive spectral features in the NIR (see Figure \ref{contrasts}). For this analysis, we exploited water and CO bands in the infrared, similarly to how visible-wavelength features such as TiO lines are used for earlier-type stars \citep[e.g.][]{mirtorabi2003,oneal2004}.

In this study, we adopted a single characteristic $\mu_\bullet$ value for both the occulted photosphere and the occulted active region. In reality, each occulted region encompasses a range of $\mu$ values. Using a single $\mu$ value should be a good approximation in most situations, given other uncertainties in the analysis. Nevertheless, in the case of giant starspots or starspot groups, this approximation might not be the most suitable.

Our study aims at assessing \textit{JWST}'s capabilities in a wide range of observational settings and scenarios. In our formalism for modelling starspot occultations during transits, we simplified the transit three-dimensional (3D) geometry in a two-dimensional analytical formulation. This treatment is insensitive to the degeneracies between starspot size, location on stellar disc, and inclination of planetary orbit, but not on the correlation between $T_\bullet$ and the starspot size. Thanks to the width of the occultation bumps, we were however able to break the starspot temperature and size degeneracy in the largest SNR cases. We also remark that our formalism to derive starspot contrast spectra is valid as long as the planet disc is completely enclosed in the stellar disc. Its extension to the transit edges is beyond the scope of this paper.

The formalism we adopted for the determination of the starspot contrast from $T_\bullet$ requires the measurement of the transit depth (Equation \ref{deltaf_f}). In the first place, this means that the uncertainty on the results depends on the precision on the transit depth measure. Additionally, as atmospheric absorption by the planet can affect this quantity, our method can be directly applied to planets with a thin atmosphere, or with a flat transmission spectrum. In other cases, the transit depth variation due to the planet's atmosphere can be represented by a wavelength-dependent $\beta (\lambda)$ factor; alternatively, our model has to be combined with a transmission spectrum retrieval. This exploration requires a specific development, which is not the focus of this study.

To date, very few M-type transiting exoplanet host stars are known with mag$_K < 8$.\footnote{Source: exoplanet.eu.}. However, the precision we achieved on starspot temperatures on M dwarfs with NIRCam/F150W2 + F322W2 is particularly relevant for measuring temperature contrasts for this stellar type, which are expected to be smaller than in FGK stars \citep{berdyugina2005,herbst2021}. Our results are promising for large contrast cases, and could be extended to large starspots, which were observed both for M \citep{berdyugina2011} and for K stars \citep[e.g.][]{morris2017}.

The simulations were performed for NIRSpec/Prism and NIRCam/F150W2 + F322W2, the modes that provide the broadest NIR wavelength coverage on \textit{JWST}. With these configurations, valuable information on the activity level of stellar hosts will come for free in the case of a stellar active feature occultation. This will possibly act as an additional constraint on the level of contamination of planetary transmission spectra: information on the average contrast of stellar active features could be used to determine the most suitable stellar models to be used in atmospheric retrievals \citep{iyer2020}. Additional information could then be gained by combining this information with the output of ground-based photometric monitoring campaigns \citep[e.g.][]{rosich2020,guilluy2020}. 

In this work, we demonstrated the feasibility of constraining the temperature of occulted starspots with no additional observation other than those used to perform transmission spectroscopy. Despite the focus on \textit{JWST}, the same method presented here could be used for any low-resolution facility working with transmission spectroscopy, such as the Atmospheric Remote-sensing Infrared Exoplanet Large-survey (\textit{Ariel}, \citealp{tinetti2018}). However, dedicated simulations should be used to model each instrument's properties.

\section*{Acknowledgements}
GB acknowledges support from CHEOPS ASI-INAF agreement n. 2019-29-HH.0. GM and GC acknowledge the support of the ARIEL ASI-INAF agreement 2021-5-HH.0. This research has made use of the SVO Filter Profile Service (http://svo2.cab.inta-csic.es/theory/fps/) supported from the Spanish MINECO through grant AYA2017-84089. 

\section*{Data availability}
The specific intensity models used for this study can be downloaded from Zenodo \citep{zenodo}, DOI 10.5281/zenodo.5609422 (https://zenodo.org/record/5609422). The code underlying this article will be shared on reasonable request to the corresponding author.
The throughput files used for our simulations can be downloaded from the SVO Filter Profile Serive (tps://jwst-docs.stsci.edu/near-infrared-camera/nircam-instrumentation/nircam-filters).  
Repositories for the software used in this study can be found in the referenced literature.

%%%%%%%%%%%%%%%%%%%%%%%%%%%%%%%%%%%%%%%%%%%%%%%%%%



\bibliographystyle{mnras}
\bibliography{biblio} % if your bibtex file is called example.bib

%\begin{thebibliography}{99}
%\end{thebibliography}

%%%%%%%%%%%%%%%%%%%%%%%%%%%%%%%%%%%%%%%%%%%%%%%%%%

%%%%%%%%%%%%%%%%% APPENDICES %%%%%%%%%%%%%%%%%%%%%

\appendix

\section{Posterior distributions for a low-occultation SNR case}\label{appendix_posterior_lowsnr}
We here report the corner plot for the low-SNR, white-light transit fit in the scenario with a 14.5 $K$ mag, M star observed with NIRSpec/Prism and a $T_\bullet = 3200$ K occulted starspot, with $\theta=0^\circ$, discussed in Section \ref{lowsnr}. In particular, it can be observed that the posterior distribution for the occultation flatness $n$ does not provide enough constraints on such parameter. As a result, a high $\beta_\mathrm{min}$ is produced, and a warm starspot solution is forced.

\begin{figure*}
\includegraphics[scale=0.25]{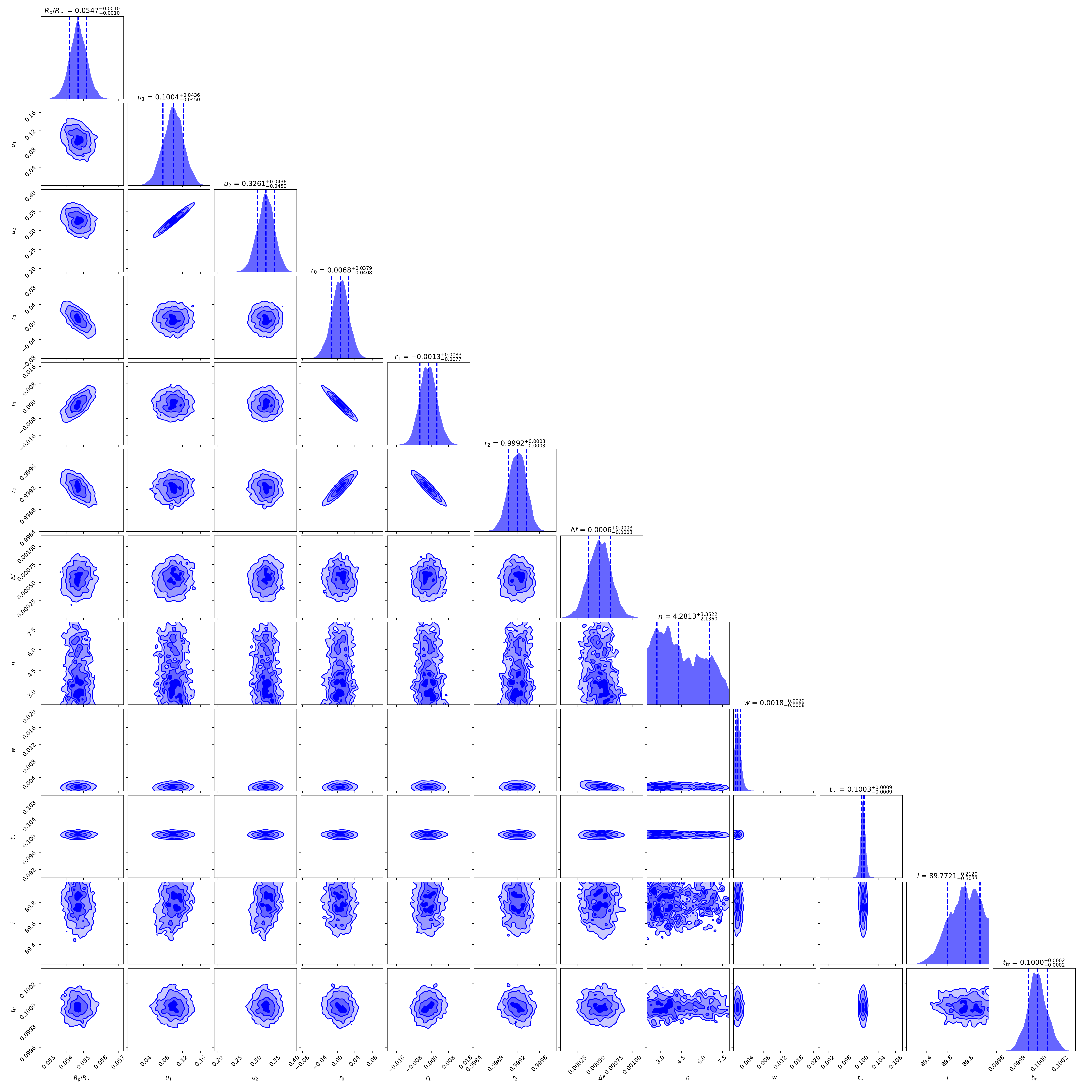}
\caption{Marginalised posterior distributions for the low-SNR, white-light transit nested sampling on a 14.5 $K$ mag, M star observed with NIRSpec/Prism and a $T_\bullet = 3200$ K occulted starspot, with $\theta=0^\circ$, as discussed in Section \ref{lowsnr}. The columns represent, from left to right: planet-to-star radius ratio, linear and quadratic limb darkening coefficient, out-of-transit quadratic trend parameters ($r_0, r_1, r_2$), flux bump $\Delta f$, occultation flatness $n$, spot width $w$, occultation time $t_\bullet$, planet orbital inclination, and transit mid-time. On large-SNR occultation cases, the $n$ parameter is tightly constrained.}
\label{corner_transit}
\end{figure*}

\section{Posterior distributions for all simulations}\label{appendix_posteriors}
We here provide a version of Figures \ref{resnirspec}, \ref{resnircam}, and \ref{reslat} which includes the posterior distributions for all scenarios. The SNR of a given solution represents the baseline for the horizontal histogram of the respective posterior. These plots allow the identification of trends in the bimodality of the solutions, as well as the significance of the 50\% percentile values represented in the first version of the Figures.

\begin{figure*}
\includegraphics[width=\columnwidth]{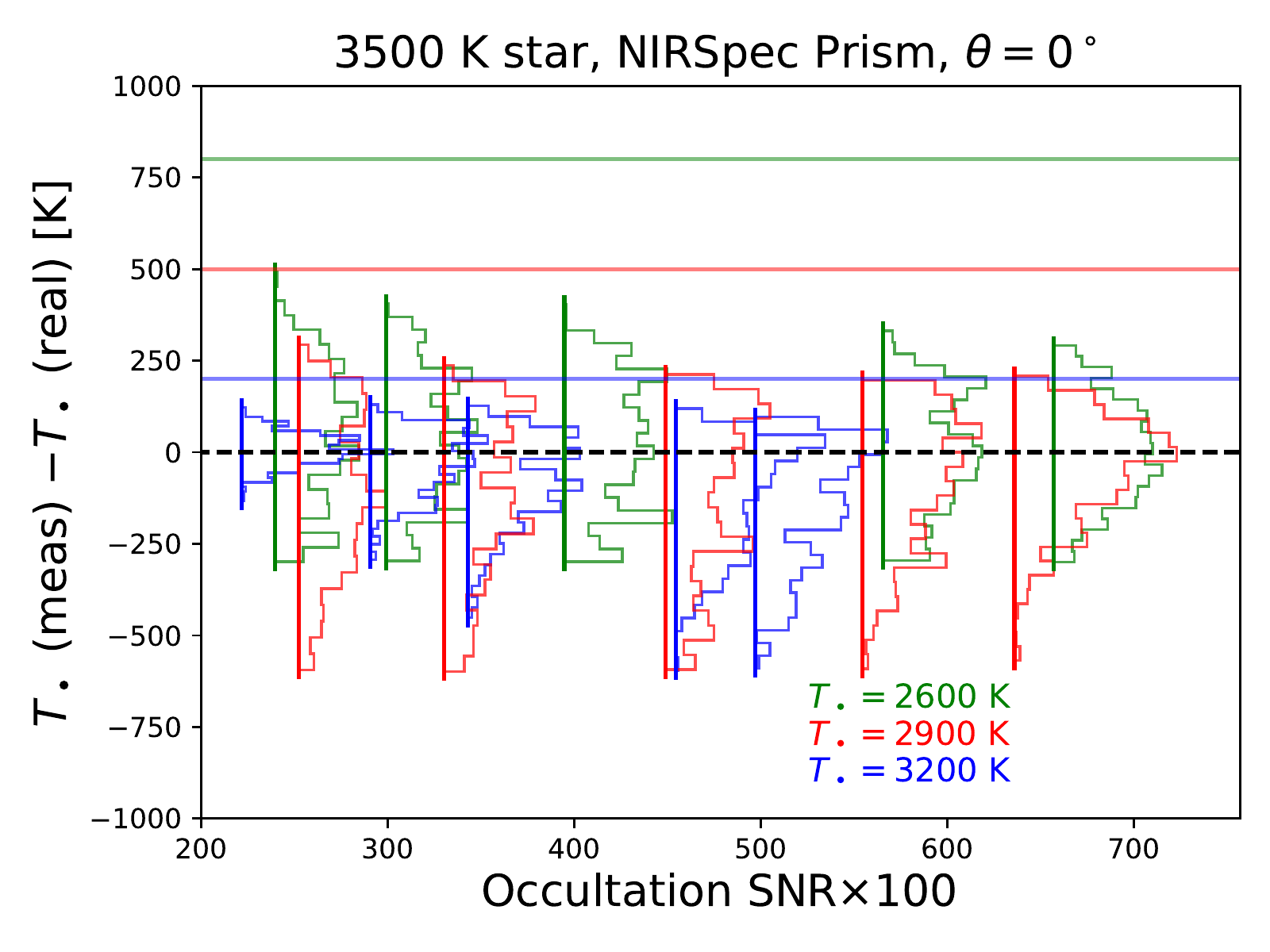}
\includegraphics[width=\columnwidth]{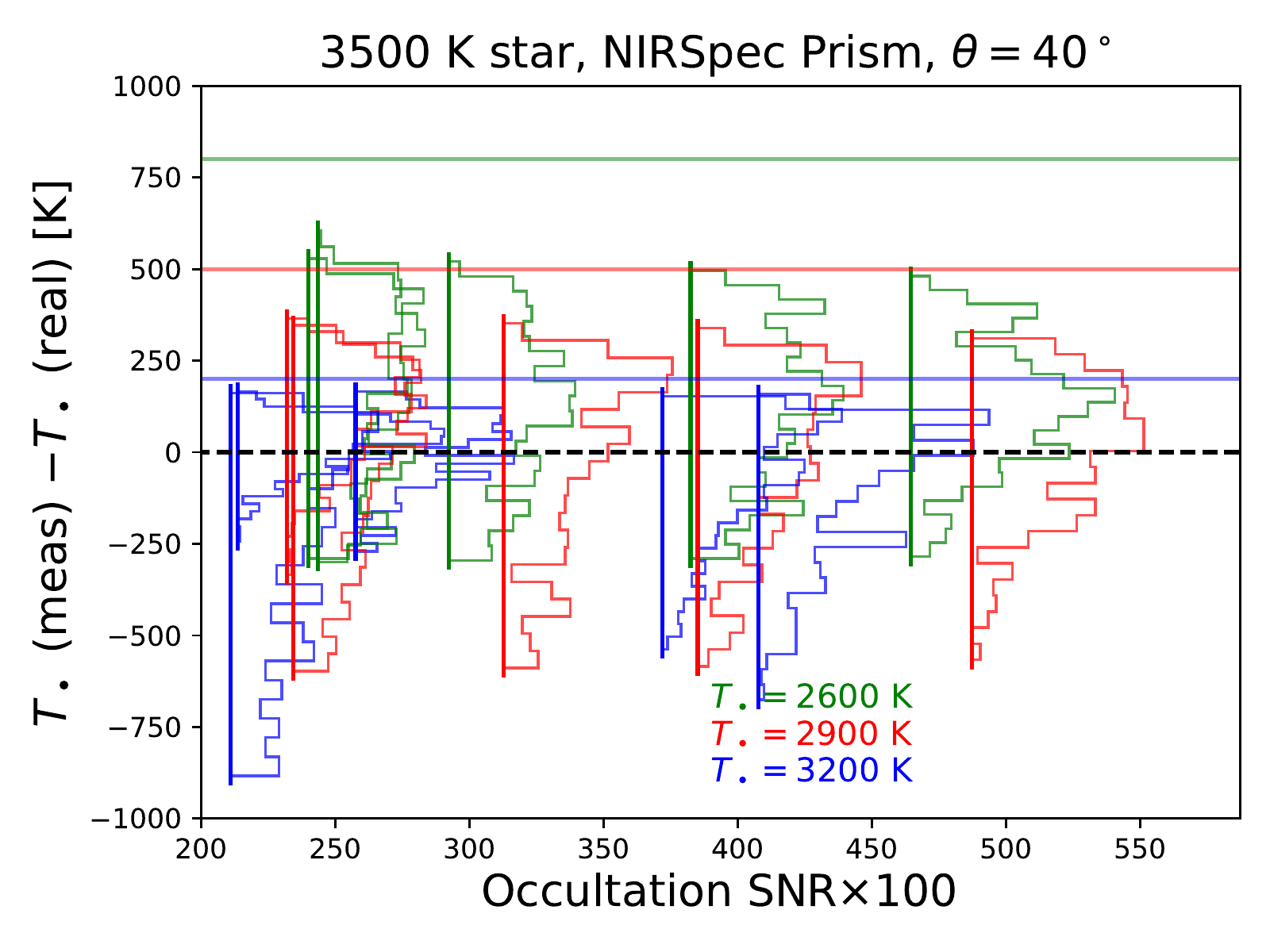}
\includegraphics[width=\columnwidth]{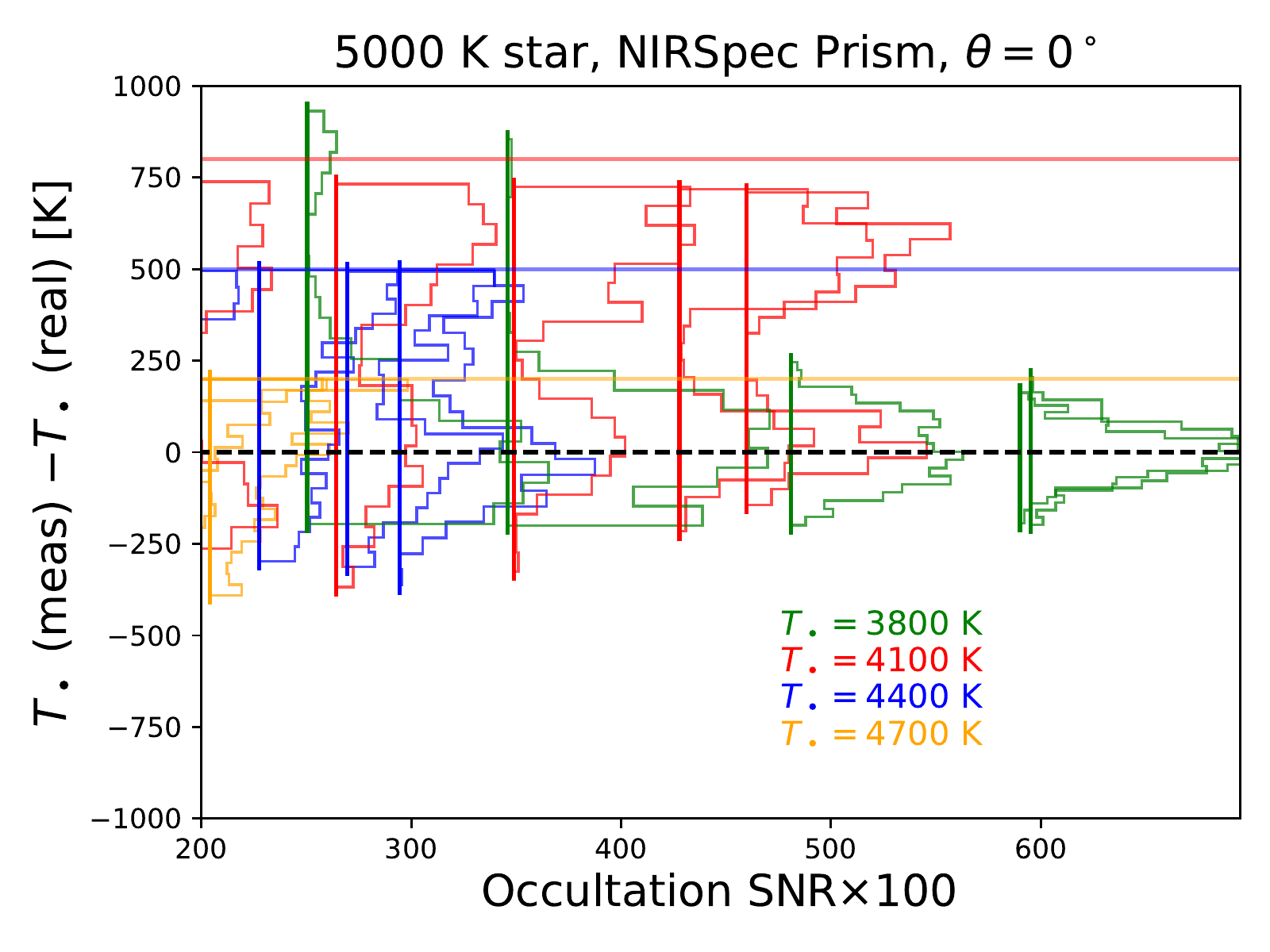}
\includegraphics[width=\columnwidth]{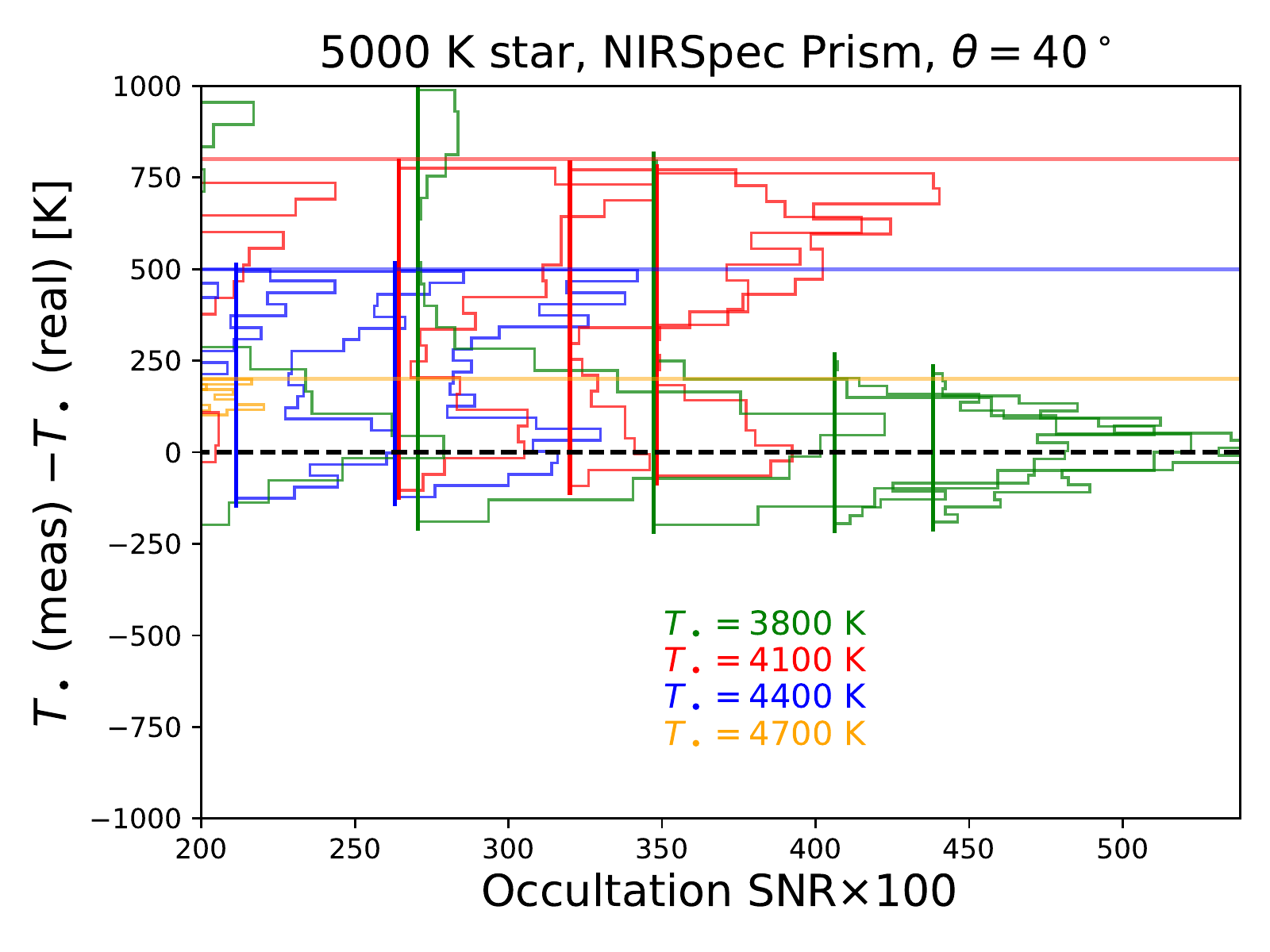}
\caption{Posterior distributions of the difference between the measured and the true $T_\bullet$ for the simulations carried out for NIRSpec/Prism, as a function of the occultation SNR multiplied for visualisation purposes. The Figures are divided by stellar temperature (rows) and starspot limb-angle (column). The baseline of each horizontal histogram indicates the SNR of the corresponding scenario and the probability in the distributions increases along the $x$-axis (different probability values in a given histogram are all related to the same SNR value). Different colours represent different true $T_\bullet$ cases, and stellar $T_\star - 100$~K values (the lowest allowed in our $T_\bullet$ fit) are marked with horizontal lines with the corresponding colour for each scenario. Bimodal distributions for the K star, 4100~K starspot, mid- to low-SNR case can be clearly distinguished; in the lowest SNR cases, only a lower limit can be assigned to $T_\bullet$.}
\label{resnirspec_histo}
\end{figure*}

\begin{figure*}
\includegraphics[width=\columnwidth]{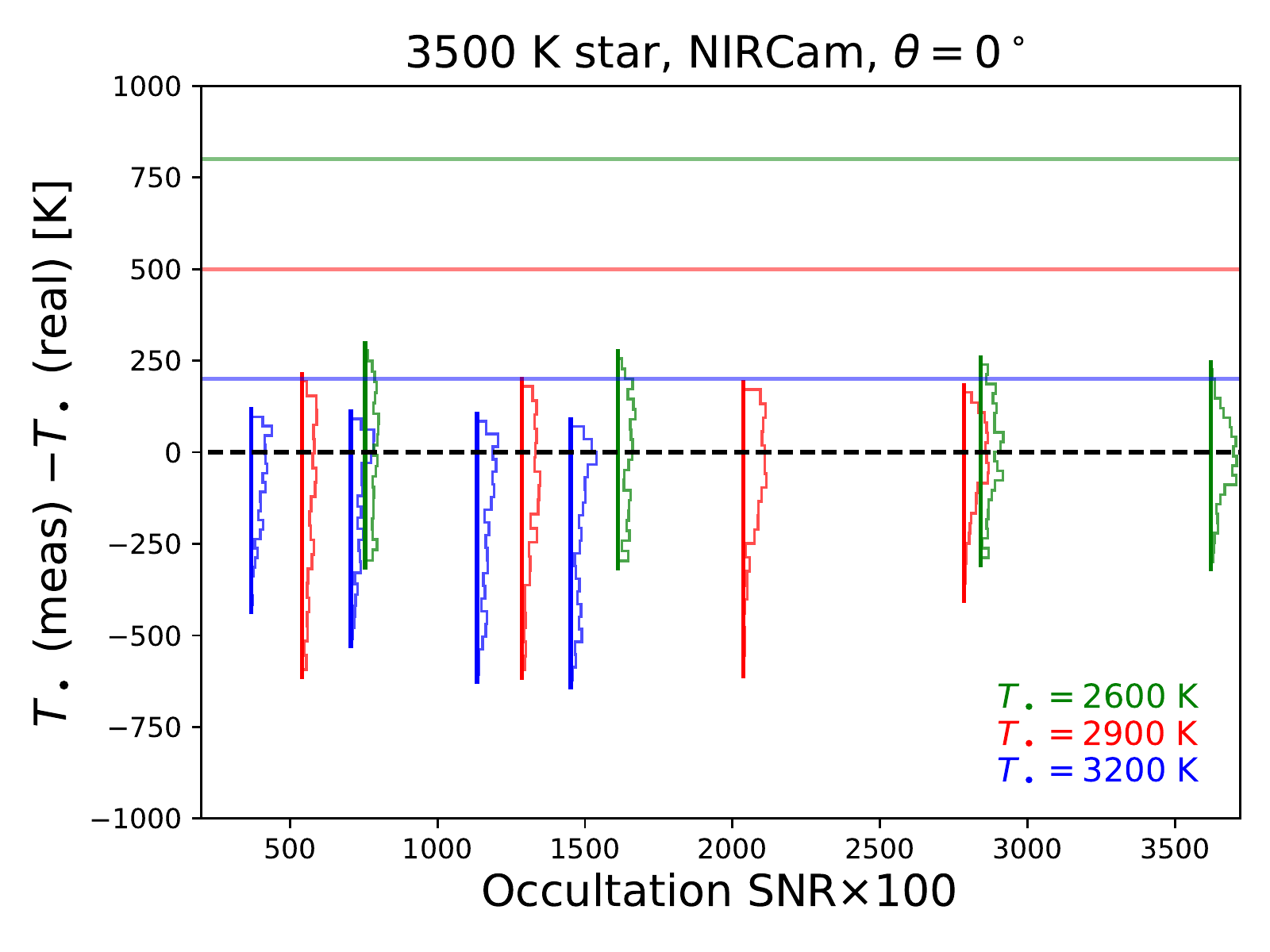}
\includegraphics[width=\columnwidth]{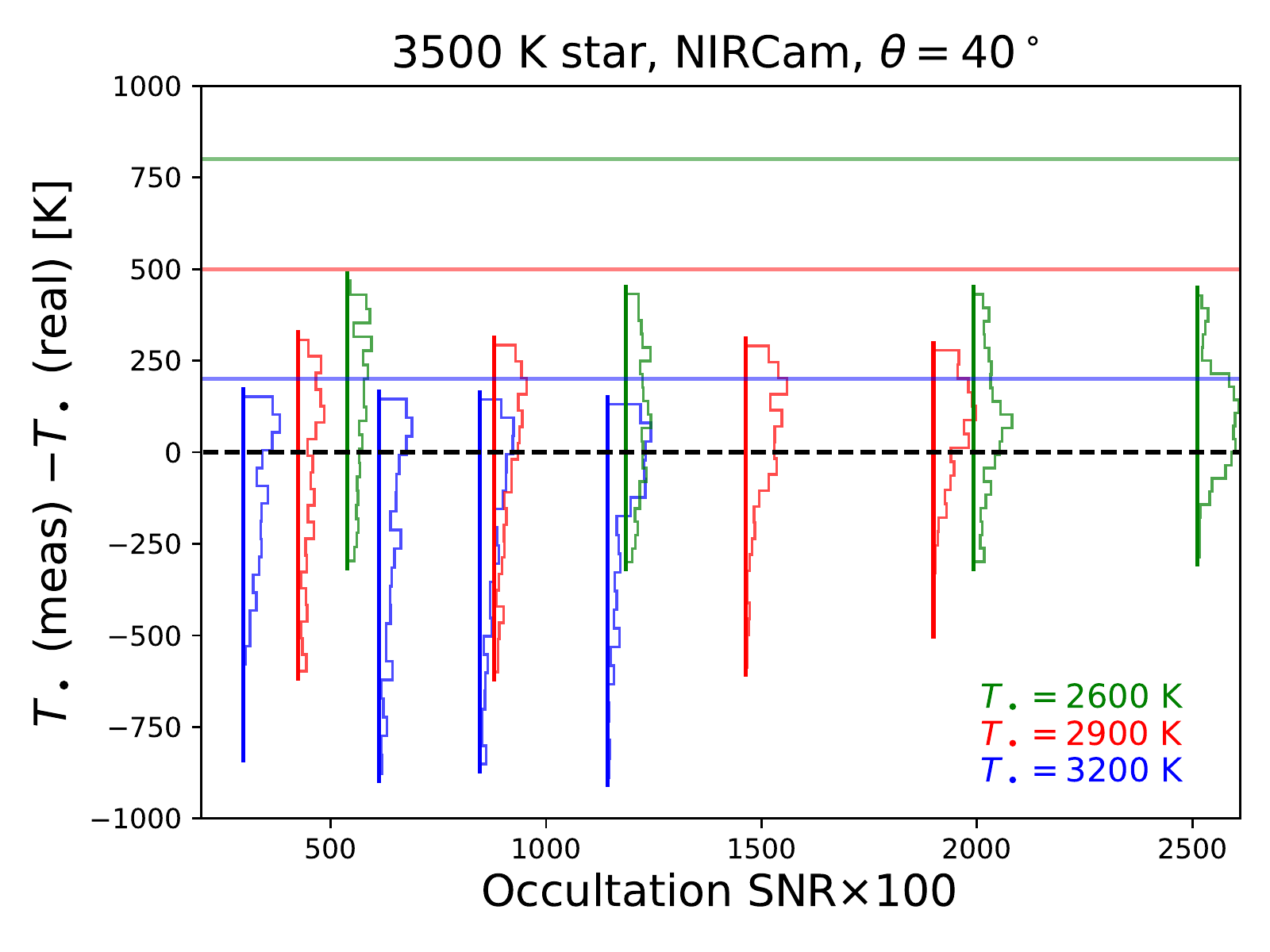}
\includegraphics[width=\columnwidth]{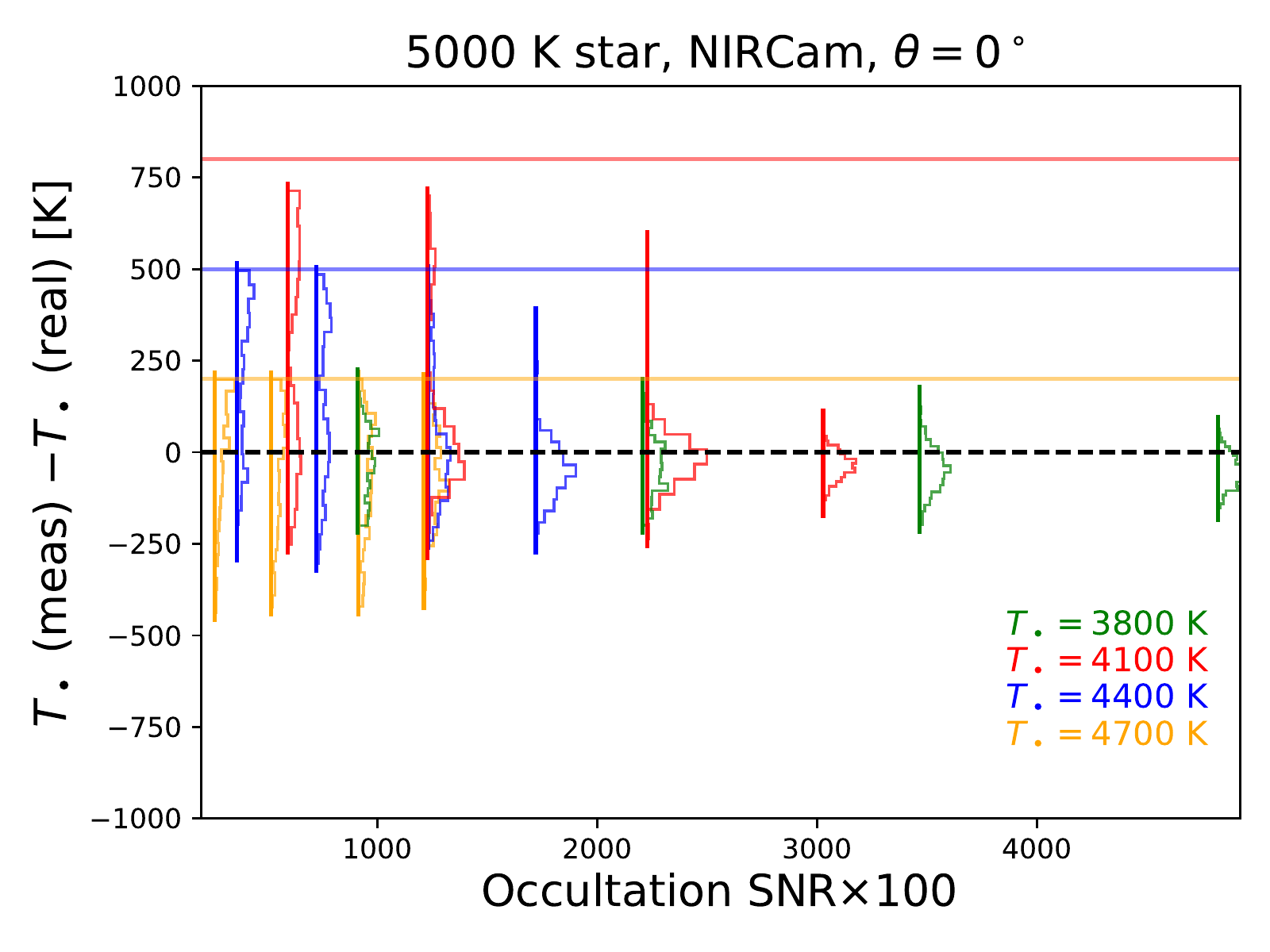}
\includegraphics[width=\columnwidth]{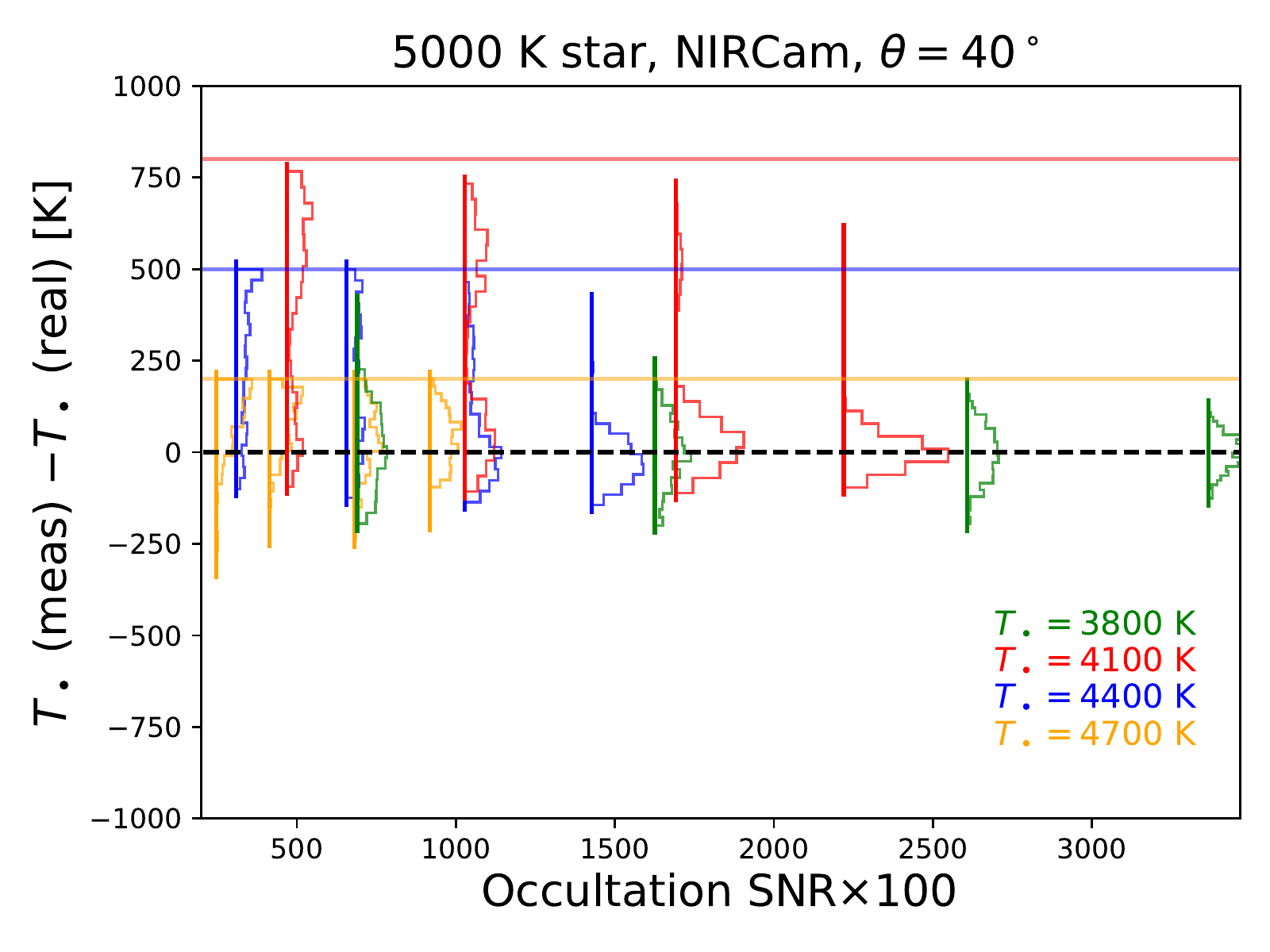}
\caption{Same as Figure \ref{resnirspec_histo}, but for NIRCam/F150W2 + F322W2.}
\label{resnircam_histo}
\end{figure*}

\begin{figure*}
\includegraphics[width=\columnwidth]{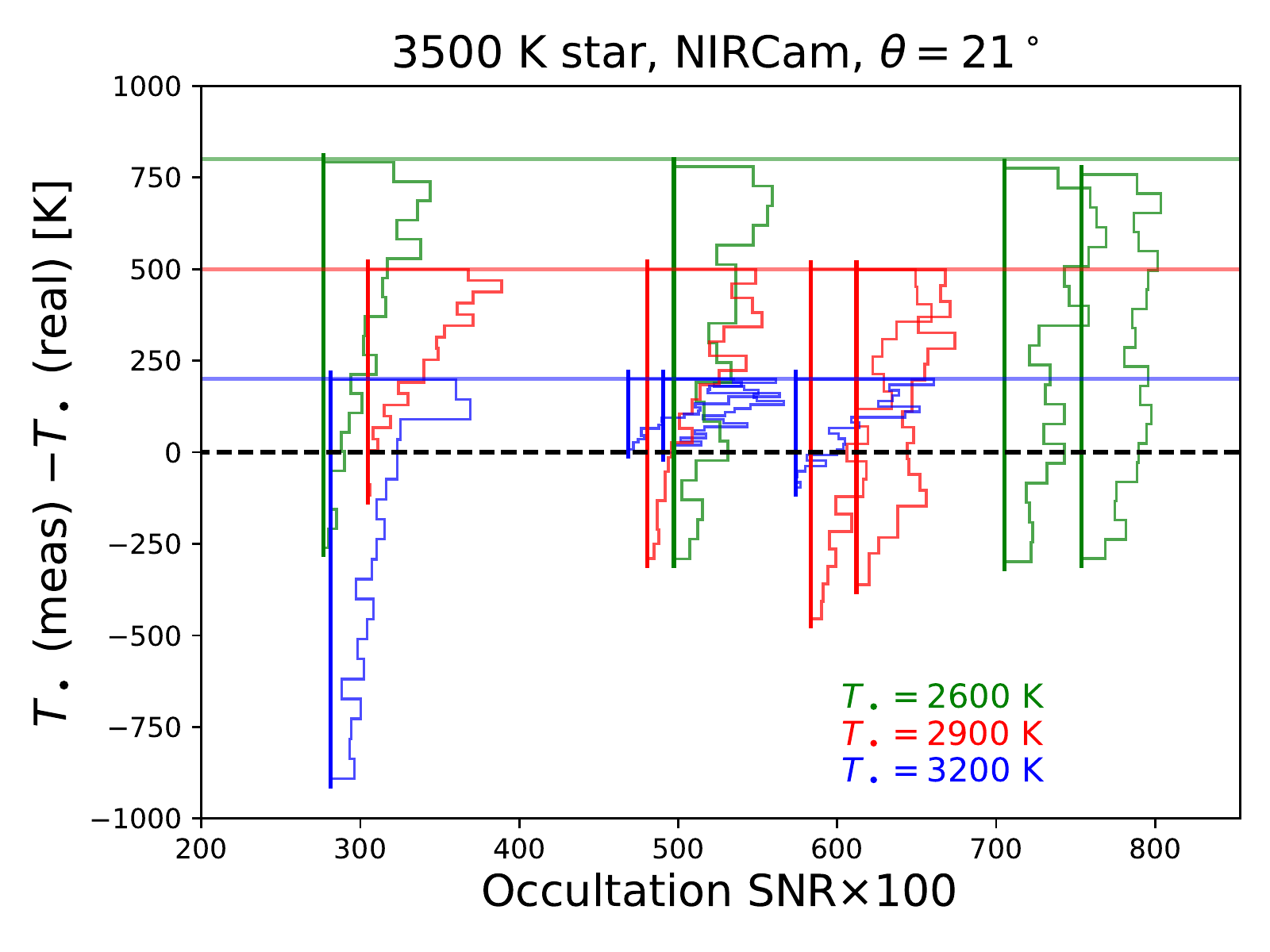}
\includegraphics[width=\columnwidth]{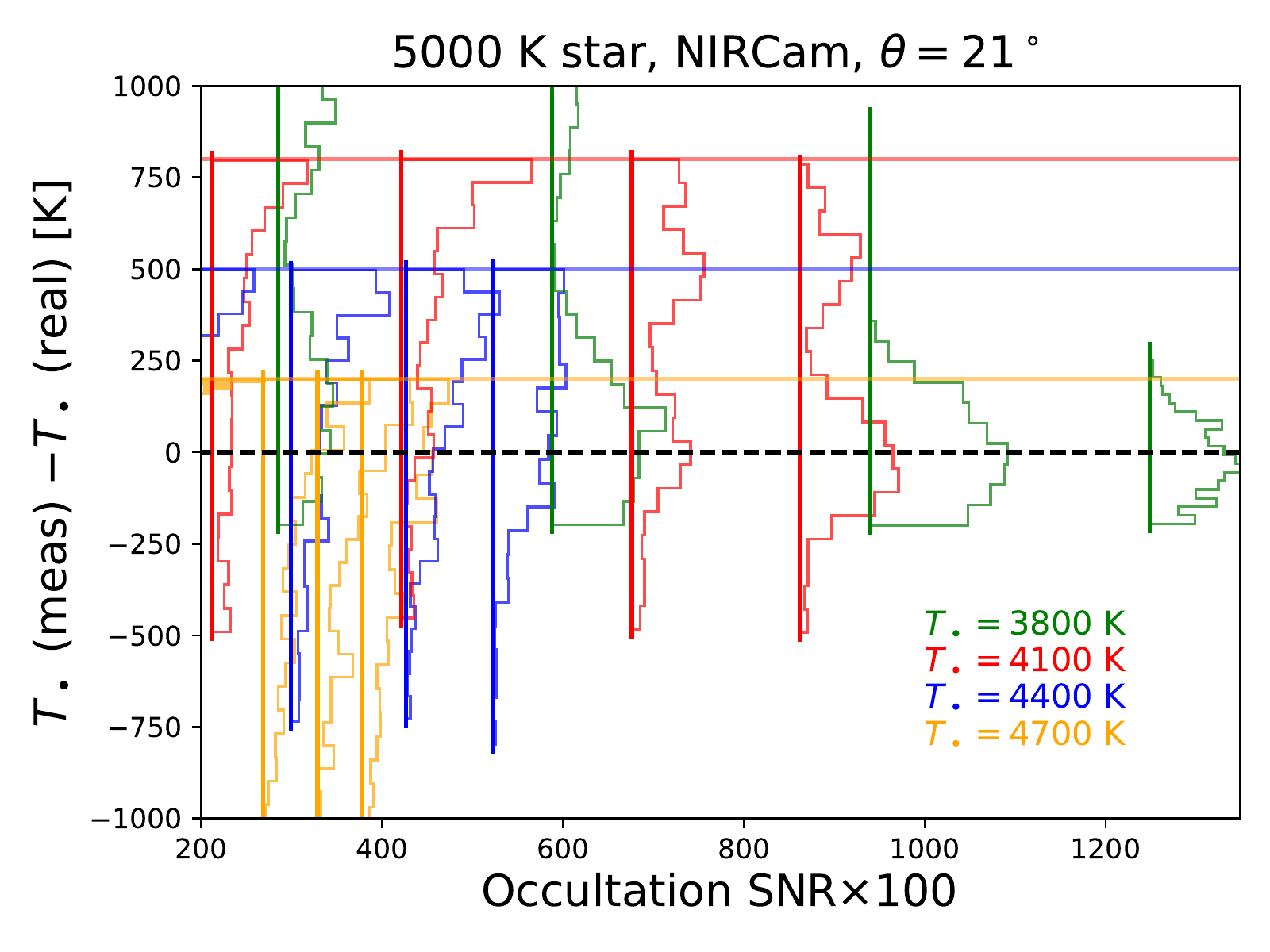}
\caption{Same as Figure \ref{resnirspec_histo} for spots at $21^\circ$ stellar latitude, for the M (left) and K (right) scenarios observed by NIRCam/F150W2 + F322W2.}
\label{reslat_histo}
\end{figure*}

%%%%%%%%%%%%%%%%%%%%%%%%%%%%%%%%%%%%%%%%%%%%%%%%%%

% Don't change these lines
\bsp	% typesetting comment
\label{lastpage}
\end{document}